\documentclass[journal,twoside,web]{ieeecolor}
\usepackage{generic,cite}
\usepackage{amsmath,amssymb,amsfonts,multicol,tikz,booktabs,multirow}
\usepackage{algorithmic,graphicx,textcomp,caption,mathtools,makecell}
\usepackage[implicit = false]{hyperref}
\usepackage[font=small]{caption}
\newtheorem{definition}{Definition}
\newtheorem{theorem}{Theorem}

\newtheorem{assumption}{Assumption}
\newtheorem{remark}{Remark}
\newtheorem{corollary}{Corollary}
\DeclareMathOperator*{\minimize}{minimize}
\DeclareMathOperator*{\subject_to}{s. t. }
\def\BibTeX{{\rm B\kern-.05em{\sc i\kern-.025em b}\kern-.08em
    T\kern-.1667em\lower.7ex\hbox{E}\kern-.125emX}}
\usepackage{fancyhdr}
\usepackage{xcolor}

\fancyfootoffset{-1em}
\fancypagestyle{copyright}{
	\chead{\footnotesize This work has been submitted to the IEEE for possible publication. Copyright may be transferred without notice, after
		which this version may no longer be accessible.}
}

\begin{document}
	
\title{Gaussian Process-Based Nonlinear Moving Horizon Estimation}
\author{Tobias M. Wolff, Victor G. Lopez, \emph{Member, IEEE}, and Matthias A. Müller, \emph{Senior Member, IEEE}
\thanks{This project has received funding from the European Research Council (ERC) under the European Union’s Horizon 2020 research and innovation programme (grant agreement No 948679). }
\thanks{Tobias M. Wolff, Victor G. Lopez, and Matthias M. Müller are with the Leibniz University Hannover, Institute of Automatic Control, 30167 Hannover, Germany (email: \{wolff,lopez,mueller\}@irt.uni-hannover.de)}
\thanks{}}

\maketitle
\thispagestyle{copyright}
\begin{abstract}
In this paper, we propose a novel Gaussian process-based moving horizon estimation (MHE) framework for unknown nonlinear systems. On the one hand, we approximate the system dynamics by the posterior means of the learned Gaussian processes (GPs). On the other hand, we exploit the posterior variances of the Gaussian processes to design the weighting matrices in the MHE cost function and account for the uncertainty in the learned system dynamics. The data collection and the tuning of the hyperparameters are done offline. We prove robust stability of the GP-based MHE scheme using a Lyapunov-based proof technique. Furthermore, as additional contribution, we derive a sufficient condition under which incremental input/output-to-state stability (a nonlinear detectability notion) is preserved when approximating the system dynamics using, e.g., machine learning techniques. Finally, we illustrate the performance of the GP-based MHE scheme in two simulation case studies and show how the chosen weighting matrices can lead to an improved performance compared to standard cost functions. 
\end{abstract}


\section{Introduction}
\label{sec:introduction}
\IEEEPARstart{M}{oving} horizon estimation (MHE) \cite{Rawlings2020} is an optimization-based state estimation technique. Loosely speaking, at each time instant, an optimization problem is solved to determine an optimally estimated state sequence, based on the available input/output measurements and a mathematical model of the physical system. The cost function of the optimization problem considers the magnitudes of the estimated process and measurement noise (often called stage costs) as well as the deviation between the initial element of the estimated state sequence and a prior estimate (often called prior weighting). The constraints of the optimization problem guarantee that the system dynamics are satisfied (based on the available mathematical model) and, potentially, that the estimated states and disturbances lie in some predefined sets. Finally, the state estimate at the current time instant is set to the last element of the optimal estimated state sequence. The main advantages of MHE are that (i) it is applicable to general nonlinear systems, (ii) system-inherent constraints can directly be considered in order to improve the estimation performance, and (iii) strong robust stability guarantees have been obtained in recent years, such as, e.g., guarantees which hold for relatively small horizon lengths \cite{Rawlings2020,Allan2021robust,Schiller2023}. 

MHE is crucially based on a mathematical model of the system. Typically, expert knowledge, first principles, or heuristics are used to derive such a model. These approaches can be time-consuming and expensive and will not always result in reliable models. An alternative is to \textit{learn} the mathematical model of the system by using some machine learning technique.  

Only a few works have studied the combination of MHE and machine learning techniques. In \cite{alessandri2011moving}, the authors propose to approximate the solution of the MHE optimization problem by a function that is parameterized using a single-hidden-layer feedforward neural network. This approach reduces the computational burden of the MHE scheme. Furthermore, the authors provide stability guarantees by relying on some technical assumptions. However, the MHE scheme and hence also the stability guarantees heavily rely on model knowledge. 

Another approach to combine learning techniques and MHE has been suggested in \cite{Cao2022}. The authors propose a so-called primal-dual estimation learning method. Loosely speaking, two functions are learned offline, (i) one function that generates state estimates online and (ii) another function that examines the optimality of the estimates. However, the work considers only linear systems, although MHE has been proven to be particularly powerful for nonlinear systems. 

An alternative machine learning technique is Gaussian process (GP) regression. A Gaussian process is defined as a collection of random variables, any finite number
of which follows a joint Gaussian distribution \cite[Def. 2.1]{rasmussen2006gaussian}. The key advantage of Gaussian processes (GPs) is that they provide a measure of how uncertain a predicted function output is (using the posterior variance). GPs have already been frequently used in the area of learning-based control, compare, e.g., \cite{beckers2019stable,Maiworm2021,hewing2019cautious}.

In the context of state estimation, GPs have recently been used for a joint dynamics and state estimation method \cite{buisson2021joint}, where the state estimation is performed by a high-gain observer and the dynamics estimation by GPs. Additionally, interval observers using GPs for partially unknown dynamics with stability guarantees have been developed in \cite{capone2019interval}. Other GP-based estimators were proposed in \cite{Ko2009,ko2007gp}, in particular GP-based extended/unscented Kalman filters (EKF/UKF), where the state transition function and the output map are approximated by GPs. Furthermore, GP-based assumed density filters were suggested in \cite{deisenroth2009analytic} \cite[Ch. 4]{deisenroth2010efficient}. 
The drawbacks related to these methods are two-fold. On the one hand, they estimate the states based on one-step recursions. In case perfect model knowledge is available, it is well known that considering a sequence of output measurements i.e., as done in MHE often results in a better performance \cite{Rawlings2020}. On the other hand, these methods do not allow to consider system-inherent constraints, the inclusion of which can, however, improve the estimation performance~\cite{Rawlings2020}.

Jointly using MHE and GPs to perform state estimation has already been suggested in some first application-oriented works in robotics. For instance, in \cite{tong2013gaussian,tong2013gaussian3D}, the authors suggest a state estimation framework, where GPs are used to directly estimate the states and not to approximate the system dynamics.
The major drawbacks of these works are that no theoretical stability analysis is done and that the state estimation scheme is limited to the full information estimation case i.e., using all available measurements to estimate the states, which becomes computationally intractable as $t$ increases. In \cite{choo2023data}, the authors suggest to approximate parts of the system dynamics by GPs and to build an MHE based on this hybrid model. However, once again, no stability analysis of the presented scheme has been done.

In this work, we propose a novel GP-based MHE framework, which uses GPs to approximate the state transition function and the output map. We exploit the uncertainty quantification provided by GPs to design the weighting matrices in the cost function of the MHE scheme. Furthermore, different from \cite{tong2013gaussian,tong2013gaussian3D,choo2023data}, we prove robust stability of the MHE scheme under mild assumptions. Finally, we show the performance of the GP-based MHE scheme in two benchmark examples. Additionally, we compare the performance to different standard model-based MHE schemes with perfect model knowledge and to other GP-based MHE schemes, where we also approximate the system dynamics by the posterior means of the learned GPs, but we (i) do not account for the uncertainty in the cost function or (ii) only consider one-step-ahead uncertainties in the cost function. In the two benchmark examples, we also show that our scheme outperforms alternative GP-based state estimation schemes as the GP-based EKF and the GP-based UKF. We argue that our scheme is a competitive alternative to the mentioned works in the literature, due to its simple set-up, the possibility to consider system-inherent constraints, the available (although conservative) stability guarantees, and the strong performance in our case studies. As a last contribution, we show that detectability can be preserved during the learning process as long as some technical assumptions e.g., depending on how fast the approximation error changes and related to the disturbances affecting the system are satisfied. 

Note that a preliminary version of parts of this work is available in the conference paper \cite{wolff2023robust}. Compared to \cite{wolff2023robust}, we here consider the \textit{propagated} uncertainty in the weighting matrices of the stage costs of MHE optimization problem as well as a prior weighting that depends on the uncertainty related to the prior estimate (which results in a more accurate uncertainty quantification). Moreover, we improve the stability analysis to obtain less conservative error bounds and analyze in much more detail the behavior of the GP-based MHE scheme in numerical examples. Finally, we here perform a detectability analysis as mentioned above.

This paper is organized as follows. In Section~\ref{sec:pre}, we explain the setting of this work. In the following Sections~\ref{sec:scheme} and~\ref{sec:robust:stb}, we introduce the GP-based MHE scheme and present the stability analysis, respectively. The detectability analysis is given in Section~\ref{sec:detec}. Finally, we close this paper by two illustrative examples and a conclusion in Sections~\ref{sec:numerical_example} and~\ref{sec:conclusion}, respectively.

\section{Preliminaries}
\label{sec:pre}
\subsection{Notation}
The set of integers greater than or equal to $a \in \mathbb{R}$ is denoted by $\mathbb{I}_{\geq a}$. Furthermore, the set of non-negative real numbers is written as $\mathbb{R}_{\geq 0}$. We denote the identity matrix of dimension~$n$ by $I_n$ and a diagonal matrix of dimension~$n$ with~$q_1, \dots, q_n$ on the diagonal entries by $\mathrm{diag}(q_1, \dots, q_n)$. A positive definite matrix $P$ is denoted by $P \succ 0$. For a vector $ x = \begin{bmatrix}
	x_1 & \dots & x_n
\end{bmatrix}^\top \in \mathbb{R}^n$ and a positive definite matrix $P$, the weighted vector norm is denoted by $||x||_P = \sqrt{x^\top P x}$. If $P=I$, we just write $||x||$, which denotes the Euclidean norm of the vector $x$.
The standard maximum (minimum) eigenvalue of a positive definite matrix~$P$ is denoted by $\lambda_{\max}(P)$ ($\lambda_{\min}(P)$) and the maximum generalized eigenvalue of positive definite matrices $P_1 = P_1^\top$ and $P_2 = P_2^\top$ is denoted by~$\lambda_{\max}(P_1, P_2)$, i.e., the largest scalar $\lambda$ satisfying $\det(P_1-\lambda P_2) = 0$. For two symmetric matrices $A$, $B$, $ A \preccurlyeq B$ means that $B-A$ is positive semidefinite. The (Pontryagin) set difference is denoted by $\ominus$, i.e., for two sets $\mathcal{A}, \mathcal{B}$, $\mathcal{A} \ominus \mathcal{B} \coloneqq \{a \in \mathbb{R}^n | a + b \in \mathcal{A}, \forall b \in \mathcal{B}\}$. For some scalar $a \in \mathbb{R}$, we write $\lfloor a \rfloor$ to denote the largest integer less than or equal to $a$. For two square positive definite matrices of the same dimensions $A,B$, we define
\begin{align*}
	\min(A,B) \coloneqq \begin{cases}
		A, \quad \text{if $A - B \preccurlyeq$ 0}\\
		B, \quad \text{otherwise.}
	\end{cases}
\end{align*}

\subsection{Introduction Gaussian Processes}
GPs are typically used to approximate some nonlinear function~$\bar{f}:\mathbb{R}^{n_d} \rightarrow \mathbb{R}$. They are fully defined by a mean function~$m: \mathbb{R}^{n_d} \rightarrow \mathbb{R}$ and a covariance function~$k: \mathbb{R}^{n_d} \times \mathbb{R}^{n_d} \rightarrow \mathbb{R}$ (also referred to as kernel). For some $d,d' \in \mathbb{R}^n$, we write
\begin{align*}
	\bar{f}(d) \sim \mathcal{GP}(m(d), k(d,d'))
\end{align*}
to denote that the function $\bar{f}$ is a random function described by a GP. 
\begin{assumption}
	\label{ass:kernel:continuous:diff}
	The covariance function~$k$ and the prior mean function~$m$ are continuously differentiable. 
\end{assumption}

Assumption~\ref{ass:kernel:continuous:diff} is needed in the theoretical analysis and is not restrictive since, e.g., the commonly used squared exponential kernel satisfies this assumption.

Next, we collect regression input data, which we denote by $D^{dt} = \begin{bmatrix} d_1^{dt} & \dots & d_N^{dt}\end{bmatrix}$ and regression output data $Y^{dt} = \begin{bmatrix}
	y_1^{dt} & \dots & y_N^{dt}
\end{bmatrix}^\top$, where the regression output~$y^{dt}$ is given by~$y^{dt} = \bar{f}(d^{dt}) + \varepsilon_n^{dt}$ and $\varepsilon_n^{dt} \sim \mathcal{N}(0,\sigma_{\varepsilon_n}^2)$. We condition the prior distribution on the training data resulting in the posterior mean $m_+$ and posterior variance $\sigma_+^2$.
For a given test input~$d_\ast$, these are given by \cite{rasmussen2006gaussian}
\begin{align*}
	&m_+(\bar{f}(d_\ast)|d_\ast, D^{dt},Y^{dt}) = \\ 
	& \hspace{0.3cm}k(d_\ast,D^{dt})(K(D^{dt},D^{dt}) + \sigma_\varepsilon^2I)^{-1}Y^{dt} \\
	&\sigma_+^2(\bar{f}(d_\ast)|d_\ast, D^{dt},Y^{dt}) =  \\
	&\hspace{0.3cm}k(d_\ast,d_\ast)- k(d_\ast,D^{dt}) (K(D^{dt},D^{dt}) + \sigma_\varepsilon^2I)^{-1}k(D^{dt},d_\ast),
\end{align*}
for $k(d_\ast,D^{dt}) = \begin{pmatrix}
	k(d_\ast,d_i)
\end{pmatrix}_{d_i \in D^{dt}} = k(D^{dt},d_\ast)^\top$, with~$k(d_\ast,D^{dt}) \in \mathbb{R}^{1 \times N}$, and $K(D^{dt},D^{dt}) = (k(d_i,d_j))_{d_i, d_j \in D^{dt}}$ with~$K(D^{dt},D^{dt}) \in \mathbb{R}^{N \times N}$, where~$\sigma_\varepsilon$ is a hyperparameter to be tuned. The covariance function and thus the posterior mean and the posterior variance depend on some hyperparameters. Given the input $D^{dt}$ and output data $Y^{dt}$, we optimize the hyperparameters by maximizing the log marginal likelihood, see, e.g., \cite[Eq. (2.30)]{rasmussen2006gaussian}.

\subsection{Problem setting}
\color{black}
In this work, one objective is to approximate the state transition function~$f: \mathbb{R}^n \times \mathbb{R}^m \rightarrow \mathbb{R}^n$ and the output map~$h: \mathbb{R}^n \times \mathbb{R}^m \rightarrow \mathbb{R}^p$ by GPs. We consider nonlinear systems of the following form 
\begin{subequations}
	\label{def:system}
	\begin{align}
		x(t+1) &= f(x(t),u(t)) + w(t) \\
		y(t) &= h(x(t), u(t)) + v(t)
	\end{align}
\end{subequations}
with $x,w \in \mathbb{R}^n$, $u \in \mathbb{R}^m$, and $y,v \in \mathbb{R}^p$. Furthermore, we assume that the states and inputs evolve in compact sets, i.e.,~$x(t) \in \mathbb{X} \subset \mathbb{R}^n$ and $u(t) \in \mathbb{U} \subset \mathbb{R}^m$, that the functions~$f$ and $h$ are continuous, and that the process and measurement noise are distributed according to a sub-Gaussian distribution as defined in \cite[Sec. 2]{Chowdhury2017}. Note that noise distributions bounded in $[-R,R]$ satisfy this definition \cite{Chowdhury2017}.

As mentioned above, the objective is to approximate the functions~$f$ and $h$ by GPs\footnote{If only parts of the model, i.e., parts of $f$ and $h$, are unknown, these can be approximated by a GP and all results in the following hold analogously.}. To this end, we learn $n+p$ GPs to approximate the state-space model~(\ref{def:system}), i.e., one GP per component of the functions~$f$ and~$h$, since GPs are typically defined for scalar outputs. Consequently, we denote the kernels associated to each component of the function~$f$ by~$k_{x_i}$ for $i = 1,\dots, n$ and the kernels associated to each component of the function~$h$ by~$k_{y_j}$ for $j = 1, \dots, p$. Furthermore, in our case, the regression input at time $t$ corresponds to $d(t) = \begin{bmatrix}
	x_1(t) & \dots & x_n(t)& u_1(t) & \dots & u_m(t)
\end{bmatrix}^\top$, i.e., the stacked state and control input at time step~$t$. The collected regression input data\footnote{The data used for regression could come from one long trajectory or several shorter ones. To simplify the notation, we here assume that one long trajectory has been collected.} is stored in $ D^{dt} = \begin{bmatrix}
d(0) & \dots & d(N-1)
\end{bmatrix}$. Note that the individual GPs are not conditioned on the same regression output data. For instance, the GP approximating the first component of the state transition function $f$ is conditioned on $X_1^{dt} \coloneqq \begin{bmatrix}
	x_1(1) & \dots & x_1(N)
\end{bmatrix}^\top$ and the GP approximating the first component of $h$ is conditioned on $Y_1^{dt} \coloneqq \begin{bmatrix}
y_1(0) & \dots & y_1(N-1)
\end{bmatrix}^\top$, where $x_i(j)$ and $y_i(j)$ denote the {$i$-th \nolinebreak} component of the offline collected state and output at time $j$, respectively. To simplify the notation, we write 
\begin{align*}
	\allowdisplaybreaks
	m_{+,x}(f|d(t),D^{dt}, X^{dt}) &\coloneqq \begin{pmatrix}
		m_{+, x_1}(f_1|d(t),D^{dt}, X_1^{dt}) \\
		m_{+, x_2}(f_2|d(t), D^{dt}, X_2^{dt}) \\
		\vdots \\
		m_{+, x_n}(f_n|d(t),D^{dt}, X_n^{dt}) 
	\end{pmatrix}, 
\end{align*}
to denote the stacked vector of the posterior means approximating the components of~$f$ and, similarly, $m_{+,y}(h|d(t),D^{dt}, Y^{dt})$ to denote the stacked vector of the posterior means approximating the components of~$h$. The learned system is defined as 
\begin{subequations}
	\label{def:learned_system}
	\begin{align}
		x(t+1) &= m_{+,x}(f|d(t),D^{dt},X^{dt}) + \check{w}(t) \\
		y(t) &= m_{+,y}(h|d(t),D^{dt}, Y^{dt}) + \check{v}(t)
	\end{align}
\end{subequations}
with $\check{w} \in \mathbb{R}^n$ and $\check{v} \in \mathbb{R}^p$. The original system dynamics (\ref{def:system}) are obtained for
\begin{align}
	\check{w}(t) & \coloneqq f\big(x(t), u(t)\big) - m_{+,x}\big(f|d(t),D^{dt},X^{dt}\big) + w(t) \label{def:auxiliary_w}\\
	\check{v}(t) & \coloneqq h\big(x(t), u(t)\big) - m_{+,y}\big(h|d(t),D^{dt},Y^{dt}\big) + v(t) \label{def:auxiliary_v}. 
\end{align}
In this work, we consider two different phases. First, we consider an offline phase, where we collect training data that is necessary to construct the posterior means and variances. In our case, the training data includes measurements of the inputs, outputs \textit{and} states as in many other works related to data/learning-based state estimation such as, e.g., \cite{Turan2021,Wolff2024robust,Ko2009,deisenroth2009analytic}. In addition, we perform the hyperparameter optimization in this offline phase. Second, we consider an online phase in which only measurements of the inputs and outputs but \textit{not} the states are available. To estimate the states in the online phase, we employ the GP-based MHE scheme that we introduce in Section~\ref{sec:scheme}. The assumption that state measurements are available in an offline phase might be restrictive in general, but is certainly fulfilled in applications where one can measure all the states of the system using a dedicated laboratory and/or expensive hardware as in the automotive, aviation, or shipping industry, compare the detailed discussion in \cite{Turan2021,Wolff2024robust}. One concrete application where this setting is fulfilled is autonomous driving, where additional sensors can be installed in a vehicle to measure all the states (which then corresponds to the offline phase). However, to save costs, these sensors are not installed in a series production. Hence, the states must be estimated in the online phase  \cite{Graeber2019,Ehlers2022}. We need the assumption of offline available state measurements to fix the physically meaningful state-space realization of the state estimates. If one is not interested in state estimates that correspond to a physically meaningful realization, one could potentially employ other methods that do not require state measurements as suggested in~\cite{Eleftheriadis2017} to design an MHE scheme.

\begin{remark}
	\label{rmk:measurement-noise-free-state-measurements}
	In the offline phase, we consider that the system states are affected by process noise, but not by measurement noise. That is, our state measurements are noise-free. This setting is commonly assumed in the literature \cite{Ko2009,hewing2019cautious,deisenroth2009analytic}, compare also \cite[Rmk. 9]{deisenroth2010efficient}. In a more realistic setting, one would need to consider that the state measurements in the offline phase are corrupted by some measurement noise (\textit{in addition} to the already here considered process noise), i.e., only $x_{\mathrm{noisy}}(t) = x(t) + r(t)$ can be measured (with $r$ being, e.g., some normally distributed noise). This more realistic setting leads to a substantially more involved regression problem since the offline regression inputs are only known up to some random noise, i.e., they are not known exactly. This problem is addressed from a practical point of view in \cite{ko2011learning,mchutchon2011gaussian,Wolff2024gaussian}. In \cite[Fig. 5]{ko2011learning}, it is shown that the state estimation error of a GP-based UKF is small if it is based on regression inputs that are affected by small input noise levels. In our recent work \cite{Wolff2024gaussian}, we extend the approach from \cite{mchutchon2011gaussian} to dynamical systems and showed in simulations that small input noise levels also result in small regression errors. In order to avoid unnecessary complication introduced by input noise, we here consider noise-free regression inputs and leave the case of noisy regression inputs as an interesting subject for future work. 
\end{remark}
At this point we can state the problem that we address in this work.

\textit{Problem Description:} Develop a GP-based state estimation framework for \textit{general} nonlinear systems with robust stability guarantees that takes system-inherent constraints into account and considers the inherent uncertainty quantification of GPs.

\section{GP-based MHE scheme}
\label{sec:scheme}
In this section, we start by presenting the novel GP-based MHE scheme in detail. First, we present the optimization problem of the GP-based MHE scheme. Second, we explain the design of the weighting matrices in the cost function based on the posterior variances of the GPs.
\subsection{Optimization Problem}
\label{subsec:opt_prob}
At each time $t$, given the past $M_t = \min\{t,M\}$ ($M$ being the horizon length) inputs $u(j)$ and measured outputs $y(j)$ for $j \in \mathbb{I}_{[t-M_t,t-1]}$, solve
\begin{subequations} \label{MHE_nom}
	\begin{align}
		\minimize_{\substack{\bar{x}(\cdot|t), \bar{w}(\cdot|t) \\ \bar{v}(\cdot|t)}} \hspace{0.2cm}  &J\big(\bar{x}(t-M_t|t), \bar{w}(\cdot|t), \bar{v}(\cdot|t),t\big) \label{cost_function_nom} \\
		\subject_to  \bar{x}(j+1|t) =&~m_{+,x}(f|\bar{d}(j|t), D^{dt},X^{dt}) +\bar{w}(j|t), \label{eq:mean_in_MHE_f}\\
		y(j) =&~m_{+,y}(h|\bar{d}(j|t), D^{dt},Y^{dt}) + \bar{v}(j|t), \label{eq:mean_in_MHE_h}  \\ &\hspace{3cm}\forall j \in \mathbb{I}_{[t-M_t, t-1]} \nonumber\\
		\bar{x}(j|t) &\in \mathbb{X} \quad \forall j \in \mathbb{I}_{[t-M_t, t]} \label{state_constraints}
	\end{align}
	with
	\begin{align*}
		\bar{d}(j|t) \coloneqq \begin{bmatrix}
			\bar{x}_1(j|t) &
			\dots &
			\bar{x}_n(j|t) &
			u_1(j) &
			\dots &
			u_m(j) 
		\end{bmatrix}^\top
	\end{align*}
	and
	\begin{align}
		J(\bar{x}(&t-M_t|t),\bar{w}(\cdot|t), \bar{v}(\cdot|t),t)  \nonumber\\
		\coloneqq& 2||\bar{x}(t-M_t|t) - \hat{x}(t-M_t)||_{(\Sigma_{x, \hat{d}(t-M_t-1|t-M_t)})^{-1}}^2 \eta^{M_t} \nonumber \\
		&+ \sum_{j = 1}^{M_t} 2\eta^{j-1}\Big( ||\bar{w}(t-j|t)||_{(\Sigma_{x,\bar{d}(t-j|t)})^{-1}}^2 \nonumber \\
		&\hspace{2cm}+ ||\bar{v}(t-j|t)||_{(\Sigma_{y,\bar{d}(t-j|t)})^{-1}}^2\Big). \label{eq:cost:function}
	\end{align}
\end{subequations}
The GP-based MHE scheme is composed of a cost function~(\ref{cost_function_nom}),~(\ref{eq:cost:function}) and some constraints~(\ref{eq:mean_in_MHE_f}) -~(\ref{state_constraints}). The notation~$\bar{x}(j|t)$ denotes the estimated state at time~$j$, estimated at time~$t$ and analogously for the estimated process noise and the estimated measurement noise denoted by~$\bar{w}$ and~$\bar{v}$, respectively. The notation $\bar{x}(\cdot|t)$ stands for the complete estimated state sequence at time $t$, i.e., $\bar{x}(\cdot|t) \coloneqq \begin{bmatrix}
		\bar{x}(t-M|t)^\top & \dots & \bar{x}(t|t)^\top
\end{bmatrix}^\top$ (and once again similarly for the estimated process and measurement noise sequences $\bar{w}(\cdot|t)$ and $\bar{v}(\cdot|t)$, respectively). The estimated process and measurement noise account for the noise in the online phase affecting the system \textit{and} potential approximation errors of the functions~$f$ and~$h$.

The cost function contains two parts. The first part is called prior weighting, which penalizes the difference between the initial element of the estimated state sequence~$\bar{x}(t-M_t|t)$ and the estimated state $\hat{x}(t-M_t)$ that was estimated $M_t$ steps in the past. The weighting matrix is chosen as $\Sigma_{x, \hat{d}(t-M_t-1|t-M_t)}$, which we will explain in detail in the following subsection. 

The second part is called stage cost, where we penalize the estimated process and measurement noise $\bar{w}$ and $\bar{v}$, respectively. The weighting matrices $\Sigma_{x,\bar{d}(t-j|t)}$ and $\Sigma_{y,\bar{d}(t-j|t)}$ are also explained in detail in the following subsection, compare~(\ref{def:weighting_matrix_x}) and~(\ref{def:weighting_matrix_y}), respectively. We additionally consider a discount factor $\eta$ which has a fading memory effect: the more a disturbance lies in the past, the less does it influence the cost function. In the context of MHE, discount factors have been introduced in \cite{Knufer2018} and proven to be very useful in robust stability proofs of different MHE schemes, compare, e.g., \cite{Schiller2022suboptimal,Knuefer2021,Arezki2023}. The required value of $\eta$ will be related to a detectability property of the system, compare Assumption~\ref{ass:lyap} in Section~\ref{subsec:main:result}.

By~(\ref{eq:mean_in_MHE_f}) and~(\ref{eq:mean_in_MHE_h}), we guarantee that the estimated state sequences satisfy the learned system dynamics given by the posterior means.  Inherent system constraints, such as, e.g., nonnegativity constraints of chemical concentrations can be considered in (\ref{state_constraints}). 
\begin{remark}
	\label{rmk:noise_constraints}
	Note that many model-based MHE schemes additionally consider known constraints on the estimated process and measurement noise in order to improve the estimation performance. Here, the estimated process and measurement noise account for possible approximation errors of the functions~$f$ and~$h$ (in addition to the actual noise), which might not be available in practice. Nevertheless, if such knowledge of the approximation errors is available, constraints on $\bar{w}$ and $\bar{v}$ can also be included in the GP-based MHE scheme~(\ref{MHE_nom}).
\end{remark} 

Once the optimization problem is solved, the optimal estimated state sequence is denoted by~$\hat{x}(\cdot|t)$ (similarly, we denote the optimal process and measurement noise by~$\hat{w}$ and~$\hat{v}$, respectively) and the estimated state at time $t$ is set to~$\hat{x}(t) \coloneqq \hat{x}(t|t) $. The cost of the optimal trajectory is denoted by~$J^\ast \coloneqq J(\hat{x}(t-M_t|t),\hat{w}(\cdot|t), \hat{v}(\cdot|t),t)$.

\subsection{Design of the Weighting Matrices}
\label{subsec:design:weighting_matrices}
As mentioned in the introduction, the weighting matrices in our MHE scheme (\ref{MHE_nom}) are selected such that they quantify the uncertainty of the predicted regression outputs. We now explain how such a design is achieved. In MHE, we estimate a whole state trajectory of length $M+1$. This estimation is based on $M$ one-step-ahead predictions. Loosely speaking, this means that regression outputs at one time instant are used as regression inputs at the next time instant. Since the regression outputs are uncertain, we need to consider this uncertainty in the regression input at the next time instant. 

Ideally, we would like to propagate a normally distributed regression input through the GPs. However, it is, in general, impossible to \textit{analytically} propagate uncertain regression inputs through GPs (e.g., when using a squared exponential kernel) in the sense
that we cannot compute the distribution of the regression outputs in closed form, compare \cite[Sec. 2.3.2]{deisenroth2010efficient}. Different approaches have been proposed in the literature to handle this issue. One popular approach is called ``moment matching". The idea is to approximate the unknown distribution of the propagated input by a Gaussian distribution with the same mean and variance as the true distribution \cite[Sec. 4.1]{deisenroth2013gaussian}. Another approach is to propagate the uncertainty by linearizing the posterior means which is a similar approach as in the EKF \cite{Ko2009,hewing2019cautious}\cite[Sec. 4.2]{deisenroth2013gaussian}. While being less accurate than moment matching, the linearization approach is computationally more efficient.  
Since we here already combine an optimization-based state estimation method (MHE) with a learning technique (GP regression), which are both computationally expensive, we follow the second approach. Comparing the two approaches is an interesting subject for future work. When linearizing the posterior means, the weighting matrices are defined as (compare  \cite{Ko2009,hewing2019cautious}\cite[Sec. 4.2]{deisenroth2013gaussian})
\begin{align}
	\Sigma_{x,\bar{d}(t-j|t)} &\coloneqq \min\Big\{ \Sigma_{w} +\Sigma_{x, \bar{d}(t-j|t)}^{\mathrm{os}}  \nonumber  \\
	&+ A_{\bar{d}(t-j|t)} \Sigma_{x, \bar{d}(t-j-1|t)}A_{\bar{d}(t-j|t)}^\top, \Sigma_x^{\max}\Big\}\label{def:weighting_matrix_x} \\
	\Sigma_{y,\bar{d}(t-j|t)} &\coloneqq \min \Big\{ \Sigma_{v} +\Sigma_{y, \bar{d}(t-j|t)}^{\mathrm{os}}   \nonumber \\
	&+ C_{\bar{d}(t-j|t)} \Sigma_{x, \bar{d}(t-j-1|t)}C_{\bar{d}(t-j|t)}^\top, \Sigma_y^{\max}\Big\} \label{def:weighting_matrix_y}
\end{align}
for $j = 1, \dots, M_t$ and $\Sigma_x^{\max} \succcurlyeq \Sigma_w \succ 0$ and $\Sigma_y^{\max}\succcurlyeq\Sigma_v \succ 0$, where $\Sigma_x^{\max}$ and $\Sigma_y^{\max}$ are user-defined constant matrices.
\begin{remark}
	\label{rmk:max:uncertainty:choice}
	In~(\ref{def:weighting_matrix_x}) and~(\ref{def:weighting_matrix_y}), the user-defined matrices standing for the maximal uncertainties $\Sigma_x^{\max}$ and $\Sigma_y^{\max}$ are needed for technical reasons in the proof\footnote{If we did not consider such a maximal uncertainty, the maximal propagated uncertainty would depend on the horizon length $M$ (due to the third terms in~(\ref{def:weighting_matrix_x}) and~(\ref{def:weighting_matrix_y})), meaning that it is not guaranteed that one can always find an~$M$ large enough such that $\mu^M \in [0,1)$ holds with $\mu^M$ from (\ref{eq:minimum_M}).} of Theorem~\ref{thm:MHE}. These uncertainties need to be fixed a priori. As will become clear in the proof of Theorem~\ref{thm:MHE}, the choice of the maximal uncertainties is a trade-off: if the maximal weighting matrices are set to large values, then one obtains more conservative error bounds~(\ref{thm:eq:expression}) since we consider the inverse of the maximal uncertainty $\Sigma_x^{\max}$ on the left-hand side of~(\ref{thm:eq:expression}). In turn, if the maximal weighting matrices in (\ref{def:weighting_matrix_x}) and (\ref{def:weighting_matrix_y}) are set to rather small values, the weighting matrices will often correspond to these constant weighting matrices since we consider the minimum in~(\ref{def:weighting_matrix_x}) and~(\ref{def:weighting_matrix_y}) which means that we do not fully exploit the inherent uncertainty quantification of the GPs in the cost function. 
\end{remark}
The first terms in~(\ref{def:weighting_matrix_x}) and~(\ref{def:weighting_matrix_y}) correspond to the diagonal matrices 
\begin{align}
	\Sigma_w &\coloneqq \mathrm{diag}(\sigma_{w_1}^2, \dots, \sigma_{w_n}^2) \label{def:sigma_w}\\
	\Sigma_v &\coloneqq \mathrm{diag}(\sigma_{v_1}^2, \dots, \sigma_{v_p}^2) \label{def:sigma_v}
\end{align}
where $\sigma^2_{w_1}, \dots, \sigma^2_{w_n}$ and $\sigma^2_{v_1}, \dots, \sigma^2_{v_p}$ are the noise variances in the online phase of each dimension of the functions~$f$ and~$h$. The second term corresponds to the one-step-ahead uncertainties $\Sigma_{x, \bar{d}(t-j|t)}^{\mathrm{os}}$ and $\Sigma_{y, \bar{d}(t-j|t)}^{\mathrm{os}}$ defined as
\begin{align}
	\Sigma_{x, \bar{d}(t-j|t)}^{\mathrm{os}} \coloneqq&  \mathrm{diag}\big(\sigma_{+,x_1}^{2}(f_1|\bar{d}(t-j|t), D^{dt},X_1^{dt}), \dots, \nonumber \\
	&\sigma_{+,x_n}^{2}(f_n|\bar{d}(t-j|t), D^{dt},X_n^{dt})\big)\\
	\Sigma_{y, \bar{d}(t-j|t)}^{\mathrm{os}} \coloneqq&  \mathrm{diag}\big(\sigma_{+,y_1}^{2}(h_1|\bar{d}(t-j|t),D^{dt},Y_1^{dt}), \dots, \nonumber \\
	&\sigma_{+,y_p}^{2}(h_n|\bar{d}(t-j|t),D^{dt},Y_p^{dt}) \big).
\end{align}
Finally, the third term in (\ref{def:weighting_matrix_x}) and~(\ref{def:weighting_matrix_y}) corresponds to the propagated uncertainties, i.e., accounts for the fact that the (online) regression inputs are uncertain, where
\begin{align}
	 A_{\bar{d}(t-j|t)} \coloneqq \frac{\partial m_{+,x}\big(f|(x,u), D^{dt}, X^{dt}\big)}{\partial x }\big|_{\bar{d}(t-j|t)} 
\end{align}
and
\begin{align}
	 C_{\bar{d}(t-j|t)} \coloneqq \frac{\partial m_{+,y}\big(h|(x,u), D^{dt}, X^{dt}\big)}{\partial x }\big|_{\bar{d}(t-j|t)}, \label{eq:derivatives}
\end{align}
stand for the linearized posterior means. The matrices $\Sigma_{x, \bar{d}(t-j-1|t)}$ and $\Sigma_{y, \bar{d}(t-j-1|t)}$ represent the uncertainties from the previous time instant. Note that the indices of the weighting matrices in~(\ref{def:weighting_matrix_x}) - (\ref{def:weighting_matrix_y}) correspond to the regression input. For instance, the uncertainty $\Sigma_{x, \bar{d}(t-j|t)}$ quantifies the uncertainty of our approximation of $\bar{x}(t-j+1|t)$. Furthermore, in the third term of~(\ref{def:weighting_matrix_y}), we consider $\Sigma_{x,\bar{d}(t-j-1|t)}$, but \textit{not}~$\Sigma_{y,\bar{d}(t-j-1|t)}$. This is because the uncertainty in the predictions of the (physical) output does not depend on previous regression outputs. The uncertainty $\Sigma_{x, \bar{d}(t-M_t-1|t)}$ is set to zero, because the initial element of the estimated trajectory is not obtained by propagating an uncertain regression input.


The effect of the weighting matrices (\ref{def:weighting_matrix_x}) - (\ref{def:weighting_matrix_y}) is the following: if a small (large) amount of training data is available, the uncertainty in the predictions is rather large (small). Since we consider the inverse of the uncertainty in the cost function, the induced weight on $\bar{w}$ and $\bar{v}$ is small (large). A small (large) weight implies that we allow for large (small) values of the estimated process and measurement noise to reconstruct the measured outputs. This is meaningful since the posterior mean will be a poor (good) approximation of the true unknown function. Thus, the weights in the cost function (\ref{eq:cost:function}) are adapted according to the quality of the GP approximations. The benefit of using this choice of cost function will also be illustrated in Section~\ref{sec:numerical_example}.

The uncertainty propagation is not only useful to account for the uncertainties in the predicted trajectories, but also serves as a natural choice for the weighting matrix in the prior weighting. By propagating the uncertainty in the described manner, we obtain an estimate of the uncertainty related to the current state estimate $\hat{x}(t)$, i.e., $ \Sigma_{x, \hat{d}(t-1|t)}$. Note that this state estimate becomes the prior $M$ time instants later and hence, the uncertainty of this prior is $\Sigma_{x, \hat{d}(t-M_t-1|t-M_t)}$ (which is used as weighting matrix in the prior weighting). As long as~$t<M$ (i.e., as long as not enough measurements are available to fill a complete horizon), a prior uncertainty~$\Sigma_{x,\hat{d}(-1|0)}$ must be set by the user, similar to other nonlinear state estimation techniques as the EKF.

\color{blue}

\color{black}

\section{Robust stability analysis}
\label{sec:robust:stb}
After having introduced the GP-based MHE scheme, we now analyze its robust stability properties. 
\subsection{Main result}
\label{subsec:main:result}
We here use the following definition of practical robust exponential stability.
\begin{definition}
	\label{def:pRES}
	A state estimator for system~(\ref{def:system}) is practically robustly exponentially stable (pRES) if there exist scalars $C_1, C_2, C_3 > 0$, $\lambda_1, \lambda_2,\lambda_3 \in [0,1)$, and $\alpha >0$ such that the resulting state estimates $\hat{x}(t)$ satisfy
	\begin{align}	
		\label{eq:pRES}	
		\|x&(t)-\hat{x}(t)\| \leq \max \Big\{ C_1\|x(0)-\hat{x}(0)\| \lambda_1^t, \nonumber \\ &\max_{j\in\mathbb{I}_{[0,t-1]}}C_{2}\|w(j)\| \lambda_2^{t-j-1},\max_{j\in\mathbb{I}_{[0,t-1]}}C_{3}\|v(j)\| \lambda_3^{t-j-1}, \alpha \Big\}
	\end{align}
	for all $t\in\mathbb{I}_{\geq0}$, all initial conditions $x(0),\hat{x}(0) \in \mathbb{X}$, and every trajectory $(x(t),u(t),w(t),v(t))_{t=0}^\infty$ satisfying the system dynamics~(\ref{def:system}).
\end{definition}
This definition states that the state estimation error is bounded at each time $t$ by (i) the initial state estimation error, (ii) the real process and measurement noise, and (iii) a positive constant $\alpha$. In this work, the introduction of the additional constant~$\alpha$ is used to capture the effect of potential approximation errors on the state estimates. 
 
To prove robust stability, we introduce the constant matrices~$\Sigma_{x}^{\min}$ and~$\Sigma_{y}^{\min}$ that satisfy
\begin{align}
	&(\Sigma_{x}^{\max})^{-1} \preceq (\Sigma_{x,\bar{d}(t-j|t)})^{-1} \preceq (\Sigma_{x}^{\min})^{-1} \label{rel:sigma_x}\\
	&(\Sigma_{y}^{\max})^{-1} \preceq (\Sigma_{y,\bar{d}(t-j|t)})^{-1} \preceq (\Sigma_{y}^{\min})^{-1} \label{rel:sigma_y}.
\end{align}
for all $t \in \mathbb{I}_\geq 0$ and all $j=1,\dots, M$ with~$\Sigma_x^{\max}$ and~$\Sigma_y^{\max}$ from (\ref{def:weighting_matrix_x}) and (\ref{def:weighting_matrix_y}), respectively. The matrices $\Sigma_{x}^{\min}$ and $\Sigma_{y}^{\min}$ can be set to the minimal possible uncertainties, which are~$\Sigma_{x}^{\min} \coloneqq \Sigma_w$ and $\Sigma_{y}^{\min} \coloneqq \Sigma_v$ with $\Sigma_w$ and $\Sigma_v$ as in~(\ref{def:sigma_w}) and~(\ref{def:sigma_v}), respectively. Furthermore, in the proof of Theorem~\ref{thm:MHE} below, we consider the cost, when the real (unknown) system trajectory is considered within the optimization problem. In this case, we denote the uncertainty of the predicted regression outputs related to the regression input at time $t-j$ by $\Sigma_{x,d(t-j)}$ and $\Sigma_{y,d(t-j)}$, which are defined as in~(\ref{def:weighting_matrix_x}) and~(\ref{def:weighting_matrix_y}) with $\bar{d}(t-j|t)$ replaced by $d(t-j)$. We refer the reader to Footnote~4 below concerning the initialization of these uncertainties.

To prove robust stability of the GP-based MHE scheme~(\ref{MHE_nom}), we need that the learned system~(\ref{def:learned_system}) satisfies a nonlinear detectability notion called incremental input/output-to-state stability ($\delta$-IOSS) as defined below.

\begin{definition}
	\label{def:detec:orig} A function $W_\delta:\mathbb{R}^n\times\mathbb{R}^n\rightarrow\mathbb{R}_{\geq 0}$ is a $\delta$-IOSS Lyapunov function for system (\ref{def:system}) if there exist $\alpha_1,\alpha_2\in\mathcal{K}_{\infty}$, $\sigma_{w},\sigma_{h}\in\mathcal{K}$, and $\eta\in[0,1)$ such that
	\begin{subequations}
		\label{eq:IOSS_Lyap}
		\begin{align}
			\label{eq:IOSS_Lyap_1}
			&\alpha_1(\|x-\tilde{x}\|)\leq W_\delta(x,\tilde{x}) \leq \alpha_2(\|x-\tilde{x}\|),\\	
			&W_\delta(f(x,u)+w,f(\tilde{x},u) +\tilde{w})  \nonumber \\ &\quad \leq \eta W_\delta(x,\tilde{x}) +\sigma_{w}(\|w-\tilde{w}\|)+\sigma_{h}(\|h(x,u)-h(\tilde{x},u)\|), \label{eq:IOSS_Lyap_2}
		\end{align}
	\end{subequations}
	for all $x,\tilde{x} \in \mathbb{X}, u \in \mathbb{U}$, and $w,\tilde{w} \in \mathbb{W} \subseteq \mathbb{R}^n$.
\end{definition}
Note that the existence of a $\delta$-IOSS Lyapunov function for system (\ref{def:learned_system}) is equivalent to the system being $\delta$-IOSS \cite{Allan2021nonlinear}. This nonlinear detectability notion (or similar versions of it) is not restrictive in the sense that it is both necessary and sufficient for the existence of a robustly stable full-order state observer~\cite{Knuefer2021,Allan2021nonlinear} and has been frequently used in the MHE literature to prove robust stability of various MHE schemes, see, e.g., \cite{Allan2021robust,Knufer2018,Schiller2022suboptimal}.
Here, we assume that the learned system (\ref{def:learned_system}) admits a quadratically bounded $\delta$-IOSS Lyapunov function, which is equivalent to the learned system being exponentially $\delta$-IOSS \cite{Allan2021nonlinear,Schiller2023}.
\begin{assumption}
	\label{ass:lyap}
	The system~(\ref{def:learned_system}) admits a $\delta$-IOSS Lyapunov function $W_\delta: \mathbb{R}^n \times \mathbb{R}^n \rightarrow \mathbb{R}_{\geq 0}$ with quadratic bounds and supply rate, i.e., there exist $ \eta\in[0,1),~P_1, P_2\succ 0$, $\tilde{Q}, \tilde{R} \succcurlyeq 0$ such that
	\begin{subequations}
		\begin{align}
			\label{ass:lyap:bounds}
			||\check{x} - \check{x}'||_{P_1}^2 \leq W_\delta(\check{x},\check{x}') \leq ||\check{x}-\check{x}'||_{P_2}^2, 
		\end{align}
		\begin{align}
			\label{ass:lyap:supply}
			&W_\delta \Big(m_{+,x}(f|\check{d},D^{dt},X^{dt}) + \check{w}, m_{+,x}(f|\check{d}',D^{dt},X^{dt}) + \check{w}^\prime\Big) \nonumber \\
			&\leq \eta W_\delta(\check{x}, \check{x}') + ||\check{w} - \check{w}^\prime||_{\tilde{Q}}^2\nonumber\\
			&+||m_{+,y}(h|\check{d},D^{dt},Y^{dt})-m_{+,y}(h|\check{d}',D^{dt},Y^{dt})||_{\tilde{R}}^2
		\end{align}
	\end{subequations}
	for all $(\check{x},u,\check{w}), (\check{x}', u, \check{w}')$ with $\check{x},\check{x}' \in \mathbb{X}$, $u \in \mathbb{U}$, where $\check{d} = \begin{bmatrix}
		\check{x}_1  &\dots& \check{x}_n &u_1 &\dots &u_m
	\end{bmatrix}^\top$, \linebreak $\check{d}' = \begin{bmatrix}
		\check{x}_1'  &\dots& \check{x}_n' &u_1 &\dots &u_m
	\end{bmatrix}^\top$.
\end{assumption}
In \cite{Schiller2023,Arezki2023}, LMI-based sufficient conditions have been introduced to guarantee the existence of a $\delta$-IOSS Lyapunov function implying that the underlying system is $\delta$-IOSS. These conditions are based on the differential dynamics of the system~(\ref{def:learned_system}) and require a linear output map, potentially in transformed coordinates compare the discussion below \cite[Ass. 2]{Schiller2023}.
To simplify the notation, we introduce the constants
\begin{align}
	\alpha_1^{\max} &\coloneqq \max_{\substack{x \in \mathbb{X}, u \in \mathbb{U}}} \Big\{\|f\big(x, u\big) - m_{+,x}\big(f|d,D^{dt},X^{dt}\big)\|_{(\Sigma_x^{\min})^{-1}} \Big\} \label{def:alpha1:max}  \\
	\alpha_2^{\max} &\coloneqq \max_{\substack{x \in \mathbb{X}, u \in \mathbb{U}}} \Big\{\|h\big(x, u\big) - m_{+,y}\big(h|d,D^{dt},Y^{dt}\big)\|_{(\Sigma_y^{\min})^{-1}}\Big\}  \label{def:alpha2:max}
\end{align}
and $\alpha^{\mathrm{max}} \coloneqq \max\{\alpha_1^{\max}, \alpha_2^{\max}\}$. These constants exist, since we assume that (i) the states and inputs evolve
in compact sets, (ii) the functions~$f$ and~$h$ are continuous, and (iii)
the applied kernel leads to a continuously differentiable posterior mean compare Assumption~\ref{ass:kernel:continuous:diff}. We are now ready to state the main result in this section.
\begin{theorem}
	\label{thm:MHE}
	Let Assumption~\ref{ass:lyap} hold with $P_1 = (\Sigma_x^{\max} + \varepsilon I)^{-1}$ (for some $\varepsilon \geq 0$), $P_2 = (\Sigma_x^{\max})^{-1}$, $\tilde{Q} = (\Sigma_x^{\max})^{-1}$ and $\tilde{R} = (\Sigma_y^{\max})^{-1}$ for $\Sigma_x^{\max}$ and $\Sigma_y^{\max}$ from~(\ref{def:weighting_matrix_x}) and~(\ref{def:weighting_matrix_y}), respectively. Then, there exist $\mu\in[0,1)$ and a minimal horizon length $\bar{M}$ such that for all $M\in\mathbb{I}_{\geq\bar{M}}$, the state estimation error of the GP-based MHE scheme~(\ref{MHE_nom}) is bounded for all $t \in \mathbb{I}_{\geq 0}$ by\footnote{In the proof, we recursively iterate a contraction over $M$ steps, as explained below (\ref{eq:initial_FIE}). Note that the initial uncertainty at the beginning of each interval is set to zero, compare the paragraph below~(\ref{eq:derivatives}). For instance, the interval from $\Sigma_{x,d(t-1)}$ to $\Sigma_{x,d(t-M)}$ is initialized with $\Sigma_{x,d(t-M-1)} = 0$ and the interval from $\Sigma_{x,d(t-M-1)}$ to $\Sigma_{x,d(t-2M)}$ with $\Sigma_{x,d(t-2M-1)} = 0$. In general, the weighing matrix~$\Sigma_{x,d(t-q-1)}$ is initialized with $\Sigma_{x,d(t-\lfloor{q/M\rfloor}\times M-1)} = 0$. With slight abuse of notation, we do not specify this initialization to improve the readability of the proof.}
	\begin{align}
		\|\hat{x}&(t) - x(t)\|_{(\Sigma_{x}^{\max} + \varepsilon I)^{-1}}  \nonumber \\
		&\leq\max \Bigg\{6\sqrt{\mu}^t\|\hat{x}(0) - x(0)\|_{(\Sigma_{x,\hat{d}(-1|0)})^{-1}},\nonumber \\
		&\max_{q\in\mathbb{I}_{[0,t-1]}}\left\{\frac{12}{1-\sqrt[4]{\mu}}\sqrt[4]{\mu}^q\|  w(t-q-1) \|_{(\Sigma_{x,d(t-q-1)})^{-1}}\right\},  \nonumber \\ 
		& \max_{q\in\mathbb{I}_{[0,t-1]}} \left\{\frac{12}{1-\sqrt[4]{\mu}}\sqrt[4]{\mu}^q\|v(t-q-1)\|_{(\Sigma_{y,d(t-q-1)})^{-1}}\right\},\nonumber \\
		& \hspace{1.5cm} \frac{12}{1-\sqrt[4]{\mu}} \alpha^{\max} \Bigg\}. \label{thm:eq:expression}
	\end{align}
	Consequently, the GP-based MHE~(\ref{MHE_nom}) is pRES according to Definition~\ref{def:pRES}. 
\end{theorem}
The proof of Theorem~\ref{thm:MHE} is given in Appendix~\ref{app:thm:proof}. It is constructive in the sense that an explicit estimate for the minimal required horizon length $\bar{M}$ is obtained (compare~(\ref{eq:minimum_M})).

Theorem~\ref{thm:MHE} guarantees that the state estimation error is bounded for any time instant~$t \in \mathbb{I}_{\geq 0}$ in dependence of (i) the initial state estimation error, (ii) the true process and measurement noise,~(iii) and the maximal difference between the true functions~$f$ and~$h$ and their approximating posterior means. This bound is less conservative than the bound obtained in the preliminary conference version of this work~\cite{wolff2023robust}, since the weights on the norm of $w$ and $v$ do not correspond to the inverse of the minimal possible uncertainty, but to the uncertainty related to the true system trajectory. 

\begin{remark}
	\label{rmk:more_training_data}
	Here, we comment on the case of having a larger amount of training data available. On the one hand, more training data in most cases decreases the uncertainty matrices~$\Sigma_{x,d(t-q-1)}$ and~$\Sigma_{y,d(t-q-1)}$ and hence also a smaller value of the maximum uncertainty~$\Sigma_{x}^{\max}$ can be chosen in~(\ref{def:weighting_matrix_x}) (while keeping weights that accurately reflect the uncertainty quantification), compare Remark~\ref{rmk:max:uncertainty:choice} above. If these additional data points are very informative, i.e., are from the region with the highest uncertainty, then the decrease of $\Sigma_x^{\max}$ will be larger than the one of the variable matrices $\Sigma_{x,d(t-q-1)}$ and~$\Sigma_{y,d(t-q-1)}$, which results in a better bound (\ref{thm:eq:expression}). On the other hand, if the new data is less informative in the sense that the maximum uncertainty $\Sigma_x^{\max}$ can only be slightly decreased, then this might not result in an improved (or even worse) error bound (\ref{thm:eq:expression}). The reason for this latter slightly counter-intuitive observation lies in the proof technique.
	We bound a manipulated expression based on the dissipation inequality (\ref{ass:lyap:supply}) of the $\delta$-IOSS Lyapunov function with the cost function, compare inequality~(\ref{eq:detec_optimal_cost}) in Appendix~\ref{app:thm:proof}. In the cost function, a larger weight in case of small uncertainties is meaningful, compare the reasoning at the end of Section~\ref{sec:scheme}. However, this is not the case for the final state estimation error bounds. To avoid this effect, one can overapproximate the right-hand side of~(\ref{thm:eq:expression}), by using the maximal possible weights $(\Sigma_x^{\min})^{-1}$ and $(\Sigma_y^{\min})^{-1}$ as weights of $w$ and $v$, respectively. Note that this more conservative bound for the right-hand side of (\ref{thm:eq:expression}) is independent of the number of training data, which means that the resulting bound (\ref{thm:eq:expression}) now can only improve with more training data due to the fact that $\Sigma_x^{\max}$ decreases. Developing error bounds that inherently exploit the uncertainty quantification and necessarily improve with more available data are an interesting subject for future work.
\end{remark}

\begin{remark}
	\label{rmk:rescale:Lyapunov}
	Notice that, if Assumption~\ref{ass:lyap} holds, then we can always select the weighting matrices $P_1 = (\Sigma_x^{\max} + \varepsilon I)^{-1}$ (for some $\varepsilon \geq 0$), $P_2 = (\Sigma_x^{\max})^{-1}$, $\tilde{Q} = (\Sigma_x^{\max})^{-1}$ and $\tilde{R} = (\Sigma_y^{\max})^{-1}$ as required in Theorem~\ref{thm:MHE} without loss of generality. This is due to the fact that one can rescale the $\delta$-IOSS Lyapunov function\cite[Rmk. 1]{Schiller2023}. 
\end{remark}

\subsection{Application of GP uniform error bounds}
So far, the error bound in Theorem~\ref{thm:MHE} depends on the maximal approximation error, captured by the constant~$\alpha^{\max}$. The dependency on the maximal approximation error can be avoided by considering uniform error bounds for GPs. These error bounds typically bound (probabilistically) the norm of the difference between some true function and the related posterior mean \cite{srinivas2012information,Chowdhury2017,fiedler2021practical,Lederer2019}. Typically, the upper bound consists of a constant multiplied by the posterior variance. 

These bounds were developed for GPs with a single output. If we learn multiple (independent) GPs for the different components of the functions~$f$ and $h$, we can apply the GP uniform error bounds separately for each component. 

Note that noise distributions bounded in $[-R,R]$ satisfy this definition \cite{Chowdhury2017}. We follow the frequentist approach, since it allows to bound the regression error for larger classes of noise. Consequently, we need the following assumption.
\begin{assumption}
	\label{ass:RKHS}
	Each component of the unknown functions~$f$ and~$h$ is a function from the RKHS\footnote{We denote the RKHS of the kernel $k$ by $\mathcal{H}_k$ and the corresponding norm by $||\cdot||_k$.} of the kernel used to approximate this component, i.e., $f_i \in \mathcal{H}_{k_{x_i}}$ and $h_j \in \mathcal{H}_{k_{y_j}}$ for all $i = 1, \dots, n$ and $j = 1, \dots, p$. Furthermore, there exist constants $B_{x_i}$ and $B_{y_j}$ such that $f_i$ and $h_j$ satisfy $||f_i||_{k_{x_i}} \leq B_{x_i}$ and $||h_j||_{k_{y_j}} \leq B_{y_j}$ for all $i = 1, \dots, n$ and $j = 1, \dots, p$. The noise affecting the regression output measurements is $R$-sub-Gaussian.
\end{assumption}

To simplify the notation, we define
\begin{align}
	\allowdisplaybreaks
	&\Delta^x \coloneqq \sqrt{\lambda_{\max}\big((\Sigma_x^{\min})^{-1}\big)} \nonumber  \\
	&\times \sum_{i = 1}^{n} \max_{x \in \mathbb{X}, \: u \in \mathbb{U}}\Big\{B_{x_i}\sigma_{+,x_i}(f|d, D^d,X_i^d) + \eta_{N,x_i}(d)\Big\} 
\end{align}
and similarly for $\Delta^y$ with $\eta_{N,x_i}$ and $\eta_{N, y_j}$ for $i = 1, \dots, n$ and $j = 1, \dots, p$ as defined in \cite{fiedler2021practical}.
Additionally, we define $\Delta^{\max} \coloneqq \max\{\Delta^x, \Delta^y\}$.

\begin{corollary}
	\label{cor:prob:stab}
	Let Assumptions~\ref{ass:lyap} and \ref{ass:RKHS} hold. Then, there exist $\mu \in [0,1)$ and a minimal horizon length $\bar{M}$ such that for all $M \in \mathbb{I}_{\geq \bar{M}}$ the state estimation error for the GP-based MHE scheme~(\ref{MHE_nom}) is bounded by 
	\begin{align}
		P\Bigg(\|&\hat{x}(t) - x(t)\|_{(\Sigma_x^{\max} + \varepsilon I)^{-1}}  \nonumber \\
		&\leq\max \Bigg\{6\sqrt{\mu}^t\|\hat{x}(0) - x(0)\|_{(\Sigma_{x,\hat{d}(-1|0)})^{-1}},\nonumber \\
		&\max_{q\in\mathbb{I}_{[0,t-1]}}\left\{\frac{12}{1-\sqrt[4]{\mu}}\sqrt[4]{\mu}^q\|  w(t-q-1) \|_{(\Sigma_{x,d(t-q-1)})^{-1}}\right\},  \nonumber \\ 
		& \max_{q\in\mathbb{I}_{[0,t-1]}} \left\{\frac{12}{1-\sqrt[4]{\mu}}\sqrt[4]{\mu}^q\|v(t-q-1)\|_{(\Sigma_{y,d(t-q-1)})^{-1}}\right\},\nonumber \\
		& \hspace{1cm} \frac{12}{1-\sqrt[4]{\mu}} \Delta^{\max} \Bigg\} \quad \forall t \in \mathbb{I}_{\geq 0}\Bigg) \geq 1-(n+p) \delta. \label{cor:prob:stab:expression}
	\end{align}
\end{corollary} 
The proof of Corollary~\ref{cor:prob:stab} is given in Appendix~\ref{app:cor:proof}. The key idea of the proof is to bound probabilistically the norm of the difference between the true function components and the posterior means by applying \cite[Prop. 2]{fiedler2021practical}. Consequently, the final state estimation error bounds are probabilistic in nature.

\section{Detectability Analysis}
\label{sec:detec}

After having presented the GP-based MHE scheme and the related stability guarantees, we address an important question regarding the use of a learning-based model as in~(\ref{def:learned_system}) in a state estimation framework. Assumption~\ref{ass:lyap}, required in our main theoretical results above, states that the learned system (\ref{def:learned_system}) is detectable in the sense that it admits a $\delta$-IOSS Lyapunov function. Clearly, if the model~(\ref{def:learned_system}) is not detectable, then its use for state estimation is not justified. Thus, it is of our interest to determine the conditions under which the learned model~(\ref{def:learned_system}) preserves the detectability property of the true system~(\ref{def:system}). Although we have focused so far on the use of GPs for function approximation, the results of this section are applicable if the state transition function $f$ and the output map~$h$ are approximated by using any machine learning technique. Furthermore, we consider $\delta$-IOSS as nonlinear detectability notion in its general form, as defined in Definition~\ref{def:detec:orig}.

As mentioned above, we assume in this section that the considered unknown system is detectable in the sense that a $\delta$-IOSS Lyapunov function exists.
\begin{assumption}
	 \label{ass:detec:orig}
	 System~(\ref{def:system}) admits a $\delta$-IOSS Lyapunov function according to Definition~\ref{def:detec:orig}.
\end{assumption}
Moreover, we define the approximation errors
\begin{align}
	e_x(x,u) &\coloneqq f(x,u)- m_{+,x}(f|(x,u), D^{dt}, X^{dt}) \label{def:ex}\\
	e_y(x,u) &\coloneqq h(x,u)- m_{+,y}(h|(x,u), D^{dt}, Y^{dt}) \label{def:ey}
\end{align}
with~$f$, $h$ from (\ref{def:system}) and $m_{+,x}$, $m_{+,y}$ from (\ref{def:learned_system}). We introduce the following additional assumption.
\begin{assumption}
	\label{ass:detec:inequality}
	Consider systems~(\ref{def:system}) and~(\ref{def:learned_system}), and Definition~\ref{def:detec:orig}. Let $\tilde{\eta}$ be such that $0 \leq  \eta  \leq \tilde{\eta}< 1$. The following inequality 
	\begin{align}
		&\sigma_w(2||e_x(x,u) - e_x(\tilde{x},u)||) \nonumber \\ 
		 &+\sigma_h(2||e_y(x,u) - e_y(\tilde{x},u)||)\leq (\tilde{\eta} - \eta) W_\delta (x,\tilde{x}) \label{eq:ass:detec:inequality}
	\end{align}
	 holds for all $x,\tilde{x} \in \mathbb{X}, u \in \mathbb{U}$.
\end{assumption}
The left-hand side of the inequality corresponds to the sum of two $\mathcal{K}$ functions, where the arguments are the norms of the difference of the state and output approximation errors at the points~$x$ and~$\tilde{x}$. This inequality will be needed in Theorem~\ref{thm:detec:orig:learned} below and quantifies how fast the approximation error is allowed to change such that detectability of the learned system is still guaranteed. In particular, for larger approximation errors~$e_x$ and~$e_y$, a larger constant $\tilde{\eta}$ will be required. Note that if the model is learned exactly, i.e., $f \equiv m_x$  and $h \equiv m_y$ (from (\ref{def:system}) and (\ref{def:learned_system})), the left-hand side of Assumption~\ref{ass:detec:inequality} equals zero, meaning that we can choose $\tilde{\eta} = \eta$. 

Moreover, we introduce the set 
\begin{align}
	\mathbb{E}_x \coloneqq &\big\{ e \in \mathbb{R}^n | e =  m_{+,x}((x,u)|D^{dt}, Y^{dt}) - f(x,u), \nonumber \\
	&\hspace{4.5cm} x \in \mathbb{X}, u \in \mathbb{U}\big\}
	\label{def:set:Ex}
\end{align}
and assume the following.
\begin{assumption}
	\label{ass:detec:sets}
	Consider the set $\mathbb{E}_x$ as defined in~(\ref{def:set:Ex}) and the set~$\mathbb{W}$ from Definition~\ref{def:detec:orig}. These sets satisfy $\mathbb{E}_x \subset \mathbb{W}$.
\end{assumption}
This assumption is required in the following theorem to guarantee that the set $\mathbb{W}$ (composed of disturbances $w$ such that there exists a $\delta$-IOSS Lyapunov function as in Definition~\ref{def:detec:orig}) captures possible approximation errors. This means that the approximation error cannot be arbitrarily large and is limited by the set of disturbances for which we can guarantee detectability. We now state the main result of this section.

\begin{theorem}
 	\label{thm:detec:orig:learned}
 	Let Assumptions~\ref{ass:detec:orig},~\ref{ass:detec:inequality}, and~\ref{ass:detec:sets} hold and consider~$w \in \mathbb{W} \ominus \mathbb{E}_x$ Then, the learned system~(\ref{def:learned_system}) admits a $\delta$-IOSS Lyapunov function.
\end{theorem}
The proof of Theorem~\ref{thm:detec:orig:learned} is given in Appendix~\ref{app:detec:proof}.

Intuitively, this theorem states that we do not lose detectability, when approximating the system dynamics as long as Assumptions \ref{ass:detec:inequality} - \ref{ass:detec:sets} are satisfied. The proof proceeds by establishing the bounds (\ref{eq:IOSS_Lyap}) for the learned system (\ref{def:learned_system}), but with the in general larger discount factor $\tilde{\eta}$ instead of $\eta$ in (\ref{eq:IOSS_Lyap_2}). This is to be expected, since the approximation error needs to be compensated. Moreover, the set to which the disturbances~$w$ can belong according to Theorem~\ref{thm:detec:orig:learned} is smaller for the learned system dynamics compared to the true dynamics (in order to admit a $\delta$-IOSS Lyapunov function), which is also due to the fact that we need to account for the approximation errors.

It is worth mentioning that Theorem~\ref{thm:detec:orig:learned} does not make use of the properties of GPs. Hence, it is a general result and also holds for other approximating techniques such as, e.g., neural networks. Moreover, here we use the general definition of a $\delta$-IOSS Lyapunov function. In case the true system satisfies the stronger property of exponential $\delta$-IOSS \cite{Schiller2023}, a similar result can be derived. 

We close this section by noting that the practical application of this theorem is, in general, limited, due to the fact that Assumptions~\ref{ass:detec:inequality} and~\ref{ass:detec:sets} are hard to verify (since, e.g., in Assumption~\ref{ass:detec:inequality}, we cannot use standard GP-based error bounds to bound the \textit{difference} of the approximation errors). In general, the state transition function and the output map are unknown and hence the approximation errors (\ref{def:ex}) - (\ref{def:ey}) cannot be computed analytically. Hence, the theorem must be interpreted as an existence result, showing that if the true (unknown) system is detectable, then the learned system preserves this property if the approximation error is small enough. Finally, as already discussed above, in practice a direct computation of a $\delta$-IOSS Lyapunov function for the learned system can be done by the methods introduced in \cite[Sec. IV]{Schiller2023} \cite{Arezki2023}.

\section{Application to batch reactor systems}
\label{sec:numerical_example}
In this section, we illustrate the performance of the GP-based MHE scheme\footnote{The implementations of all the state estimation schemes are publicly available at: \url{https://doi.org/10.25835/yc7erc4p}}. The scheme is implemented using the software Casadi \cite{andersson2019casadi} and the nonlinear program solver IPOPT \cite{wachter2006implementation}. In the first four subsections, we consider the following Euler-discretized system 
\begin{align}
	\label{example_system}
	x_1(t+1) &= x_1(t) + T(-2k_1 x_1^2(t) + 2k_2x_2(t)) + w_1(t) \nonumber \\
	x_2(t+1) &= x_2(t) + T(k_1x_1^2(t) - k_2x_2(t)) + w_2(t) \\
	y(t) &= x_1(t) + x_2(t) + v(t) \nonumber 
\end{align}
with sampling time $T = 0.1$, constants $k_1 = 0.16$, $k_2 = 0.0064$ which corresponds to a batch reactor system \cite[Ch. 4]{Rawlings2020}, \cite{Tenny2002}. This system is a benchmark example in the MHE literature, since other nonlinear state estimation techniques, such as the EKF, can fail to converge or produce physically implausible state estimates, compare \cite{Rawlings2020}. 

\subsection{GP-based MHE scheme}
\begin{figure}[t!]
	\centering
	\includegraphics[width= 0.48\textwidth]{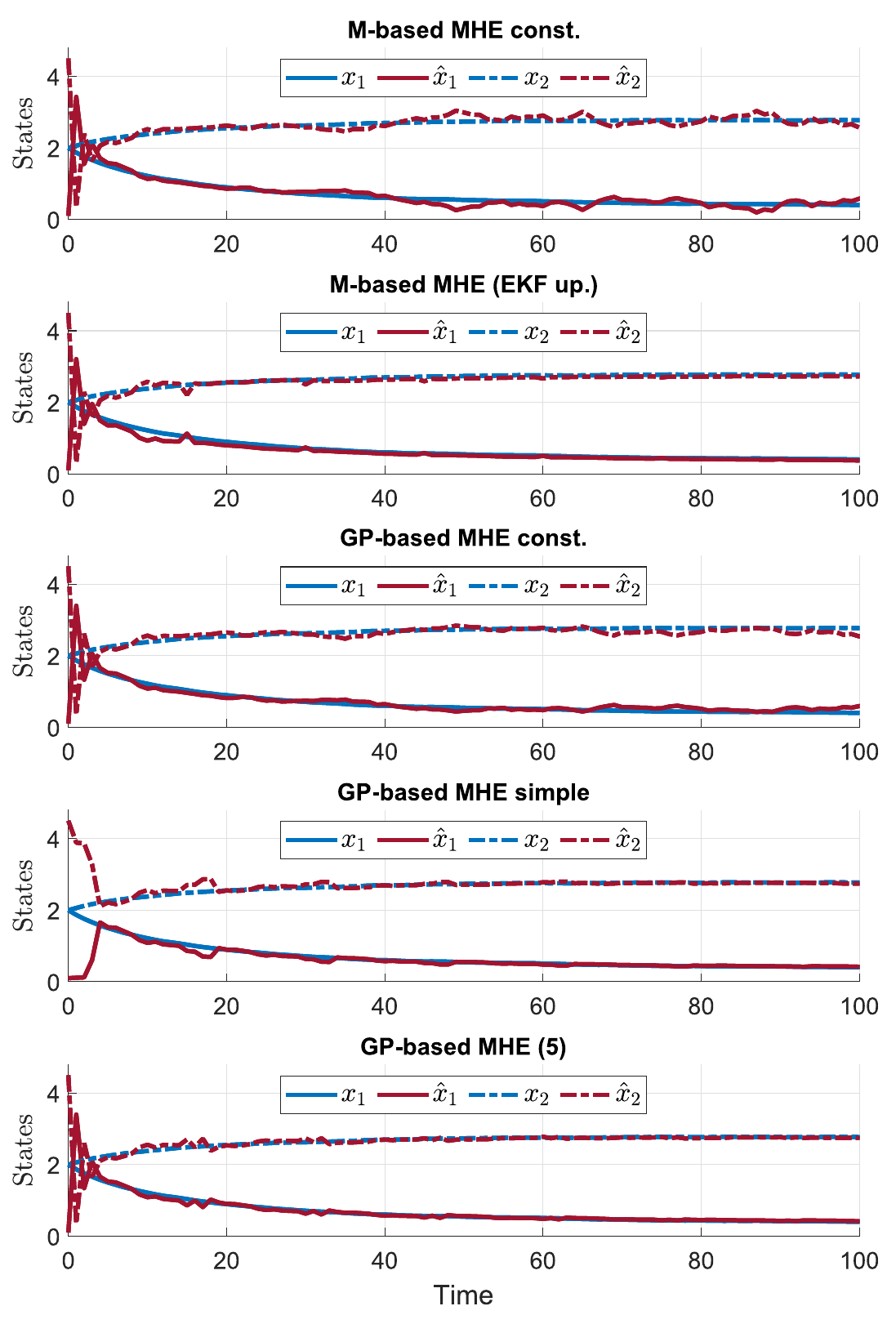}
	\caption{Simulation examples of the implemented MHE schemes for system~(\ref{example_system}). In the first plot (top), the performance of a model-based MHE scheme as proposed in~\cite{Schiller2023} is illustrated (called ``M-based MHE const."). In the following plot (called ``M-based EKF"), we illustrate the performance of a model-based MHE scheme similar to \cite{Schiller2023}, but with an EKF update in the prior weighting, as proposed in \cite{Tenny2002}. In the third plot, we show the performance of a GP-based MHE scheme similar to~(\ref{MHE_nom}), but with constant $P$, $Q$, and $R$ matrices (denoted by ``GP-based MHE const."). In the fourth plot, we illustrate the performance of the GP-based MHE scheme (\ref{eq:cost:function:const}) (called ``GP-based MHE simple"), which does not use propagated uncertainties, but only one-step-ahead uncertainties. In the last plot, we illustrate the performance of the GP-based MHE scheme as introduced in~(\ref{MHE_nom}) (denoted by ``GP-based MHE (5)").}
	\label{fig:GP:based:MHE}
\end{figure}

As mentioned in Section~\ref{sec:pre}, we here consider two different phases. In the offline phase, we assume to have measurements of input/output/state trajectories available. This data is generated by simulating 31 time steps for three different initial conditions ($x_{0,1} = \begin{pmatrix} 3 & 1 \end{pmatrix}^\top$, $x_{0,2} = \begin{pmatrix} 1 &3 \end{pmatrix}^\top$, and $x_{0,3} = \begin{pmatrix} 2 & 4 \end{pmatrix}^\top$) and normally distributed noise with mean $\mu_w = 0$ and variance $\Sigma_w = 10^{-5} I_n$ concerning the process noise affecting the states and $\mu_v = 0$ and $\Sigma_{v} = 10^{-3}I_p$ concerning the noise affecting the output measurements. Using this dataset, we optimize the hyperparameters by maximizing the log marginal likelihood. Then, we switch to the online phase and apply the GP-based MHE scheme~(\ref{MHE_nom}). We consider $M = 15$, $\eta = 0.91$, $\hat{x}(0) = \begin{pmatrix} 0.1 & 4.5 \end{pmatrix}^\top$, $x(0) = \begin{pmatrix} 2 & 2 \end{pmatrix}^\top$ and normally distributed measurement noise with mean $\mu_v = 0$ and variance $\Sigma_{v} = 10^{-3}I_p$, as well as normally distributed process noise with mean $\mu_w  = 0$ and variance $\Sigma_{w} = 10^{-5}I_n$. The initial weighting matrix of the prior is selected as $\Sigma_{\hat{d}(-1|0)} =  P^{-1}$ where $P$ is the prior matrix from \cite[Sec. V. A]{Schiller2023}. In additional simulations with different values of~$P$, we observed that varying the value of $P$ does not crucially influence the performance of the GP-based MHE scheme, compare Table~\ref{tab:prior_uncertainty} below. The maximal uncertainties are chosen as $\Sigma_{x}^{\max} = \mathrm{diag}(10^5, 10^5)$ and $\Sigma_{y}^{\max} = 10^5$. We also analyzed the influence of different maximal uncertainties, see Table~\ref{tab:max:uncertainty} below. We know that the states evolve in a compact set $\mathbb{X} = \{x\in \mathbb{R}^2| 0.1 \leq x_i \leq 4.5, \: i =\{1,2\}\}$. We implement the weighting matrices~(\ref{def:weighting_matrix_x}) and~(\ref{def:weighting_matrix_y}) by computing an upper bound of the maximal eigenvalue of the propagated uncertainty which can be done by using Gershgorin's circle theorem. Alternatively, one could compute the eigenvalues symbolically, which however only works for square matrices of dimensions three (or less) in Casadi. In the simulation example, we observed that the upper bound of the propagated uncertainty, i.e., $\Sigma_x^{\max}$ was never used. Hence, for practical application, one could also omit the implementation of the $\min$ operation and only consider the first terms in~(\ref{def:weighting_matrix_x}) and~(\ref{def:weighting_matrix_y}). The results of the GP-based MHE scheme are illustrated in the last panel of Figure~\ref{fig:GP:based:MHE} called ``GP-based MHE (\ref{MHE_nom})". As guaranteed by Theorem~\ref{thm:MHE}, the GP-based MHE scheme is pRES.

\begin{remark}
	\label{rmk:verification:deltaIOSS}
 	We verified Assumption~\ref{ass:lyap} as required by Theorem~\ref{thm:MHE} with $\Sigma_x^{\max}$ as specified above (by assuming knowledge of the linear output map) for $\{x \in \mathbb{R}^2| 0.5 \leq x_i \leq 4.5, i =\{1,2\}\}$. The minimal horizon length $\bar{M}$ (using the approximate values for~$\Sigma_{x}^{\max}$) provided by Theorem~\ref{thm:MHE} corresponds to $\bar{M} = 259$. The large magnitude of this value is due to some conservative steps in the proof technique (compare the steps between~(\ref{eq:afterJmax}) and (\ref{eq:initial_FIE}) in the proof of Theorem~\ref{thm:MHE} in Appendix~\ref{app:thm:proof}), in which we bound the variable weighting matrix of the prior by the maximal possible weighting matrix. The implementation of a GP-based MHE scheme with a horizon length $M =  \bar{M}$ such that robust stability is guaranteed is too computationally complex to be implemented on a standard PC with current computing capacities. Hence, in the simulation example, we choose the horizon length $M$ to achieve a good performance in practice as commonly done in the MHE literature, which was already the case for $M = 15$. 
\end{remark}

\subsection{Comparison to further GP/model-based MHE schemes}
\label{subsec:comparison:performance:example}
In this section, we compare the performance of the GP-based MHE scheme (\ref{MHE_nom}) to different GP-based and model-based MHE schemes. 

First, we consider another GP-based MHE scheme, in which we also approximate the system dynamics by the posterior means of GPs, but use constant matrices in the cost function~(\ref{eq:cost:function}) instead of the varying matrices that account for the uncertainties in the learned model (abbreviated by ``GP-based MHE const.").  Next, we consider the following GP-based MHE scheme 
\begin{align}
	J(\bar{x}(t-&M_t|t),\bar{w}(\cdot|t), \bar{v}(\cdot|t),t)  \nonumber\\
	\coloneqq& 2||\bar{x}(t-M_t|t) - \hat{x}(t-M_t)||_{(\tilde{\Sigma}_{x,\hat{d}(t-M_t -1|t-M_t)})^{-1}}^2 \eta^{M_t} \nonumber \\
	&+ \sum_{j = 1}^{M_t} 2\eta^{j-1}\Big( ||\bar{w}(t-j|t)||_{(\tilde{\Sigma}_{x,\bar{d}(t-j|t)})^{-1}}^2 \nonumber \\
	&\hspace{2cm}+ ||\bar{v}(t-j|t)||_{(\tilde{\Sigma}_{y,\bar{d}(t-j|t)})^{-1}}^2\Big) \label{eq:cost:function:const}
\end{align}
for 
\begin{align}
	\tilde{\Sigma}_{x,\bar{d}(t-j|t)} &\coloneqq \Sigma_{w} +\Sigma_{x, \bar{d}(t-j|t)}^{\mathrm{os}}  \label{def:weighting_matrix_x:const} 
\end{align}
and $\tilde{\Sigma}_{y,\bar{d}(t-j|t)}$ defined similarly. Further, we set $\tilde{\Sigma}_{x,\hat{d}(t-M_t -1|t-M_t)}$ analogously to $\Sigma_{x,\hat{d}(t-M_t -1|t-M_t)}$ in (\ref{eq:cost:function}). Compared to (\ref{def:weighting_matrix_x}) and~(\ref{def:weighting_matrix_y}), in this case we do not consider the propagated uncertainties and hence only have a rough estimate of the uncertainty within the estimated trajectories. We refer to this scheme by ``GP-based MHE simple". Furthermore, we implement a model-based MHE scheme as suggested in \cite{Schiller2023} (called ``M-based MHE const.") using perfect model knowledge. Finally, we consider another model-based MHE scheme similar to the one proposed in \cite{Schiller2023}, but with an EKF update in the prior weighting, as detailed in \cite{Tenny2002} (called ``M-based MHE (EKF up.)"), again using perfect model knowledge. To compare the schemes, we use the same horizon length $M$, the same discount factor $\eta$, the same true initial condition, the same prior estimate, the same noise realizations, and (where applicable) the same weighting matrices. The performance of the different schemes is illustrated in Figure~\ref{fig:GP:based:MHE}. It can be seen that the GP-based MHE scheme presented in this paper and the model-based MHE scheme with an EKF update show the best performance.

To compare the different schemes in more detail, we consider 100 different initial conditions (sampled uniformly from the interval $x(0) \in [1,3]^2$) and 100 different noise realizations (with the same means $\mu_w$, $\mu_v$ and variances $\Sigma_w$, $\Sigma_v$ as specified above) and simulate the five different MHE schemes as described above. Table~I shows the mean squared error (MSE) (averaged over the~100 simulations), defined as $\mathrm{MSE} \coloneqq \frac{1}{n T}\sum_{i=1}^{T} \sum_{j=1}^{n} (x_j(i)- \hat{x}_j(i))^2$, where~$T$ denotes the number of time steps, here $T = 150$, as well as the corresponding standard deviation.
Furthermore, we report the average computation times (to compute one state estimate) (and the corresponding standard deviation) in Table~I. We obtained the simulations using a standard PC (Intel(R) Core(TM) i7-10875H CPU @ 2.30GHz (16 CPUs) processor with 16 GB RAM).

From Table~I, we conclude that the GP-based MHE scheme as proposed in~(\ref{MHE_nom}) is computationally demanding. This is not surprising, since we combine two computationally demanding techniques: GPs and MHE. The computation times get considerably smaller in case we only consider the one-step-ahead uncertainties. When considering constant weighting matrices, the computation time further decreases substantially, mainly because no inversion of the weighting matrices in the cost function is necessary anymore. Moreover, this method still takes more time than the model-based MHE schemes, since the computation of the posterior mean is more demanding than the evaluation the true functions~$f$ and~$h$. 

In turn, the MSE of the proposed MHE scheme (\ref{MHE_nom}) is (much) better compared to the MSE of the GP-based MHE scheme considering constant weighing matrices and the GP-based MHE scheme (\ref{eq:cost:function:const}) considering only one-step-ahead uncertainties. This observation underlines the importance to consider the uncertainty of the learned model in the weighting matrices. In particular, this shows how the uncertainty quantification which is inherent to GP regression can beneficially be incorporated into MHE in order to improve the estimation performance. The GP-based MHE scheme~(\ref{MHE_nom}) even slightly outperforms the model-based MHE scheme proposed in \cite{Schiller2023}, where perfect model knowledge was assumed. In the model-based MHE scheme proposed in \cite{Schiller2023} (as well as in many other model-based MHE schemes in the literature), constant weighting matrices are used. This results in a trade-off: one can trust the prior ($P$), the model ($Q$), or the measurements ($R$). In the example above, we typically have a bad prior available. To compensate this bad prior, we choose a small weight for $P$ (and a rather large weight on $R$), such that the estimates converge quickly. However, in the steady state, this choice implies that we trust the noisy measurements a lot. Consequently, the state estimates fluctuate strongly as visible in plot referred to as ``M-based const." in Figure~\ref{fig:GP:based:MHE} which explains the large MSE. To counteract this observation, one can design a model-based MHE scheme with a weighting on the prior that is updated according to the EKF update rule, compare \cite{Tenny2002} for details. In that case, we quantify the uncertainty of the prior estimate and use this uncertainty in the cost function. In that case, we have initially a low weight on the prior. As soon as the estimates improve, the weighting on the prior increases, which reduces the fluctuations in the state estimates, especially in the steady state. This model-based MHE scheme outperforms all other schemes due to the described choice of the weighting matrices and due to the perfect model knowledge.

In conclusion, note that the GP-based MHE scheme (\ref{MHE_nom}), for which no traditional system identification technique based on first principles, heuristics or expert knowledge has been applied, outperforms the model-based MHE scheme without EKF update (but perfect model knowledge) and performs only slightly worse than the model-based MHE scheme with an EKF update (again considering perfect model knowledge). These results suggest that the GP-based MHE can be an interesting alternative in applications with large sampling times (and hence computation times do not play a crucial role), such as, e.g., in process or biomedical engineering. In order to ensure real-time applicability also in systems with small sampling times, different extensions will be examined in future work, compare Section~\ref{sec:conclusion}.


\begin{table}[t!]
	\label{tab:metrics_simulation}
	\centering
	\renewcommand{\arraystretch}{1.3}
	\begin{tabular}{lcccc} 
		\toprule
		& MSE & std MSE & $\bar{\tau}$ in s& std $\bar{\tau}$\\
		\midrule 
		M-based MHE const. \cite{Schiller2023} & 0.0807 &  0.0187 & 0.0246 & 0.0033 \\
		M-based MHE (EKF up.) & 0.0667 & 0.0179 & 0.0248 & 0.0029 \\
		GP-based MHE const. & 0.1628 & 0.0810 & 0.6561 &  0.0127 \\
		GP-based MHE simple (\ref{eq:cost:function:const}) & 0.0994 & 0.0413 & 0.9557 &  0.1791\\
		GP-based MHE (\ref{MHE_nom}) & 0.0690 & 0.0221 & 1.9006 & 0.3506 \\
		\bottomrule
	\end{tabular}
	\caption{Performance comparison of five different MHE schemes. We report the mean squared error and the mean computation times (to compute one state estimate) as well as the corresponding standard deviations of 100 different simulations.}
\end{table}

\begin{table}[t!]
	
	\centering
	\renewcommand{\arraystretch}{1.3}
	\begin{tabular}{lcccc} 
		\toprule
		&  MSE & std MSE& $\bar{\tau}$ in s & std $\bar{\tau}$\\
		\midrule 
		GP-based EKF \cite{Ko2009} & 0.1147 &  0.0435 & 0.0093 & 0.0005 \\
		GP-based UKF \cite{Ko2009}& 8.2193 & 14.5743 & 0.0274  & 0.0011  \\
		GP-based MHE (\ref{MHE_nom}) & 0.0690 &  0.0221 &  1.9006 &  0.3506 \\
		\bottomrule
	\end{tabular}
	\caption{Performance comparison of different GP-based state estimation schemes. We consider (i) a GP-based EKF \cite{Ko2009} (ii) a GP-based UKF \cite{Ko2009} and (iii) the GP-based MHE scheme as suggested in~(\ref{MHE_nom}). Once again, we report the mean squared error and the mean computation times as well as the corresponding standard deviations of 100 different simulations. Note that the UKF fails to converge in some simulations (i.e., in 24 simulations, the MSE is above 10). If one leaves out these cases, then the averaged MSE is 1.0374 and the related standard deviation 1.5987.}
	\label{tab:GP:scheme:comparison}
\end{table}

\subsection{Comparison to other GP-based state estimation methods}
In this section, we compare the developed GP-based MHE scheme (\ref{MHE_nom}) to other GP-based state estimation schemes. To this end, we consider a GP-based EKF and a GP-based UKF as developed in \cite{Ko2009}. To have a fair comparison, we learn one set of GPs to describe the system dynamics, which are then used for the GP-based MHE scheme, GP-based EKF, and the GP-based UKF.

The state estimation results are given in Table~\ref{tab:GP:scheme:comparison}. From this table, we conclude that the GP-based MHE scheme~(\ref{MHE_nom}) outperforms the alternative state estimation schemes in terms of the MSE. This is to be expected, since in the model-based case (i.e., considering MHE, EKF, and UKF with perfect model knowledge), it is well known that MHE outperforms EKF and UKF for batch reactor examples \cite{Rawlings2020}. In this example, the UKF
takes a long time to converge and even fails to converge for some initial conditions, which explains the rather bad performance. Furthermore, we note that the computation times are smaller for the GP-based EKF and the GP-based UKF compared to the here proposed GP-based MHE scheme (\ref{MHE_nom}). This is due to the nonlinear optimization problem that needs to be solved at every time instant. Hence, these results underpin that the here proposed GP-based MHE scheme is particularly interesting for applications with rather large sampling times, as it can outperform other GP-based state estimation techniques but comes with increased computational complexity.

\subsection{Analysis of design parameters in the GP-based MHE scheme}
Finally, in this section, we perform an analysis of the impact of the prior uncertainty $P$ and the maximal uncertainties $\Sigma_x^{\max}$ and $\Sigma_y^{\max}$ on the performance of the GP-based MHE scheme~(\ref{MHE_nom}). To this end, we consider the same setting as in Section~\ref{subsec:comparison:performance:example}. 

To test the influence of the prior uncertainty, we perform simulations considering three different values of $P$, namely $P = I_n$, $P = 10^{-1}I_n$, and $P = 10^{-2}I_n$ and report the MSE and the average computation times per simulation in Table~\ref{tab:prior_uncertainty}. As one can see from this table, the influence of the prior uncertainty is small. This is due to the fact that this prior uncertainty appears in the first term of the cost function (\ref{eq:cost:function}) as $\Sigma_{x,\hat{d}(-1|0)}= P^{-1}$ only for $t<M$, and then gets updated via (\ref{def:weighting_matrix_x}), and hence its influence decays rather quickly. The difference in the computation times is also negligible.

Next, we analyze the influence of the maximal uncertainties $\Sigma_x^{\max}$ and $\Sigma_y^{\max}$ on the performance of the GP-based MHE scheme. We tested different values as indicated in Table~\ref{tab:max:uncertainty}. From this table, we conclude that small values of the maximal uncertainty deteriorate the performance (see last row of Table~\ref{tab:max:uncertainty}), since the weighting matrices of the cost function no longer correspond to the actual uncertainty. In turn, using these small maximal uncertainties simplifies the nonlinear program of the GP-based MHE scheme, which results in a considerably smaller computational complexity. 

\subsection{Application to further batch reactor example}
In this subsection, we underpin the strong performance of the GP-based MHE scheme by considering a second batch reactor example with the following Euler-discretized dynamics \cite{Rawlings2020}
\begin{subequations}
	\label{example_system2}
\begin{align}
	x_1(t+1) =& x_1(t) + T(-k_1x_1(t) +k_{-1}x_2(t) x_3(t)) + w_1(t) \\
	x_2(t+1) =& x_2(t) + T(k_1x_1(t) -k_{-1}x_2(t) x_3(t) \nonumber \\
			 &-2k_2 x_2^2(t)+k_{-2}x_3(t)) + w_2(t)  \\
	x_3(t+1) =& x_3(t) + T(k_1x_1(t) -k_{-1}x_2(t) x_3(t) \nonumber \\
			 &+k_2 x_2^2(t)-k_{-2}x_3(t)) + w_3(t), \\
	y(t) =& R_c(x_1(t) + x_2(t) + x_3(t)),
\end{align}
\end{subequations}
with $T = 0.25$, $k_1 = 0.5$, $k_{-1} = 0.05$, $k_2 = 0.2$, $k_{-2} = 0.02$, and $R_c = 32.84$. This system is another benchmark example in the model-based case, since the model-based EKF and UKF can take a long time to converge or produce physically implausible state estimates (in this case negative state estimates which are not meaningful since the states correspond to chemical concentrations which can only be nonnegative), while MHE shows a very good performance. 

Here, we compare the performance of the developed GP-based MHE scheme~(\ref{MHE_nom}) once again to the GP-based EKF and UKF, which are state-of-art methods to perform GP-based state estimation. We consider a similar simulation setting as in the previous subsection and do not state all the simulation details for space reasons. We evaluate the schemes for 100 different noise realizations and initial conditions, while keeping the same prior estimate. 

As one can see from Table~\ref{tab:GP:scheme:comparison:second example}, the GP-based MHE scheme (while being more computationally demanding) once again outperforms the alternatives in terms of the MSE. As in the model-based case, we observe in the simulations that the GP-based EKF and the GP-based UKF (i) produce negative state estimates and (ii) take a long time to converge explaining the rather poor performance. 

\begin{table}[t!]
 	\centering
 	\renewcommand{\arraystretch}{1.3}
 	\begin{tabular}{lcccc} 
 		\toprule
 		& MSE & std MSE & $\bar{\tau}$ in s& std $\bar{\tau}$\\
 		\midrule 
 		$P = 10^{-2}I_n$ & 0.0772 & 0.0337 & 1.6930 & 0.2665 \\
 		$P = 10^{-1}I_n$ &  0.0785 & 0.0337 & 1.6469 & 0.2613 \\
 		$P = I_n$ & 0.0970 &  0.0354 & 1.6597 & 0.2740 \\
 		\bottomrule
 	\end{tabular}
 	\caption{ Performance of the GP-based MHE schemes for different values of the prior uncertainty $P$. We illustrate the MSE and the mean computation times to compute one state estimate.} 	\label{tab:prior_uncertainty}
 \end{table}

\begin{table}[t!]
	\centering
	\renewcommand{\arraystretch}{1.3}
	\begin{tabular}{lcccc} 
		\toprule
		&  MSE & std MSE & $\bar{\tau}$ in s & std $\bar{\tau}$ \\
		\midrule 
		\makecell[l]{$\Sigma_x^{\max} = 10^{-1}I_n$, \\ $\Sigma_y^{\max} = 10^{-1}I_p $} & 0.0690 & 0.0221 & 1.5479 & 0.2480 \\[0.9em]
		\makecell[l]{$\Sigma_x^{\max} = 10^{-3}I_n$, \\ $\Sigma_y^{\max} = 10^{-3}I_p $} & 0.0679 & 0.0197 & 1.5078 & 0.2732 \\[0.9em]
		\makecell[l]{$\Sigma_x^{\max} = 10^{-5}I_n$, \\ $\Sigma_y^{\max} = 10^{-5}I_p $} & 0.0816 & 0.0327 & 0.4391 & 0.0448 \\
		\bottomrule
	\end{tabular}
	\caption{Performance of the GP-based MHE scheme for different maximal weighting matrices. Once again, we report the mean squared error and the mean computation times to compute one state estimate.}	\label{tab:max:uncertainty}
\end{table}
\begin{table}[t!]
	\centering
	\renewcommand{\arraystretch}{1.3}
	\begin{tabular}{lcccc} 
		\toprule
		&  MSE & std MSE & $\bar{\tau}$ in s & std $\bar{\tau}$\\
		\midrule 
		GP-based EKF \cite{Ko2009} & 1.5795 & 0.2378 & 0.0175 & 0.0059 \\
		GP-based UKF \cite{Ko2009}& 1.8299 & 0.1751 & 0.0349  & 0.0012  \\
		GP-based MHE (\ref{MHE_nom}) & 0.4171 & 0.0884 & 2.3981 & 0.3794 \\
		\bottomrule
	\end{tabular}
	\caption{Performance comparison of different GP-based state estimation schemes for system (\ref{example_system2}).}
	\label{tab:GP:scheme:comparison:second example}
\end{table}
\color{black}

\section{Conclusion and Outlook}
\label{sec:conclusion}

In this paper, we introduced a novel GP-based MHE scheme. The scheme advances state-of-the-art MHE schemes by~(i) approximating the system dynamics using the posterior means of GPs, and~(ii) considering the uncertainty of the learned model directly in the cost function of the MHE scheme. We proved pRES of the GP-based MHE scheme based on standard assumptions. Moreover, we showed that nonlinear detectability (here $\delta$-IOSS) can be preserved for the learned system if the true (unknown) system is $\delta$-IOSS and the approximation error is small enough. We applied the GP-based scheme to batch reactor examples and showed that the proposed GP-based MHE scheme performs similarly well as model-based MHE schemes, where the model is exactly known. 

The key idea of the proposed framework is to combine GPs and MHE, which are both computationally demanding. An interesting subject for future research is whether sparse GP techniques, such as e.g., sparse GPs using pseudo inputs \cite{snelson2005} (as, for instance, the fully independent training conditional model), or further approaches to propagate the uncertainty as suggested in \cite{Lahr2023} can result in a comparable performance while reducing the computational complexity. Alternatively, one can examine how suboptimal/fast MHE approaches \cite{Schiller2022suboptimal,Kuehl2011,Alessandri2017} can be combined with the GP-based MHE scheme. Furthermore, the proposed scheme is crucially based on offline available state measurements. If one was able to relax this assumption, the results of this work could be applied to a broader range of applications.

Moreover, the current stability proof results in error bounds that are rather conservative and partially counter-intuitive (in the sense that more data does not necessarily imply smaller error bounds). Hence, an interesting subject for future work is to obtain error bounds that are less conservative and consistently improve with more available data.

\bibliography{jext_ref}
\bibliographystyle{IEEEtran}

\begin{IEEEbiography}[{\includegraphics[width=1in,height=1.5in,clip,keepaspectratio]{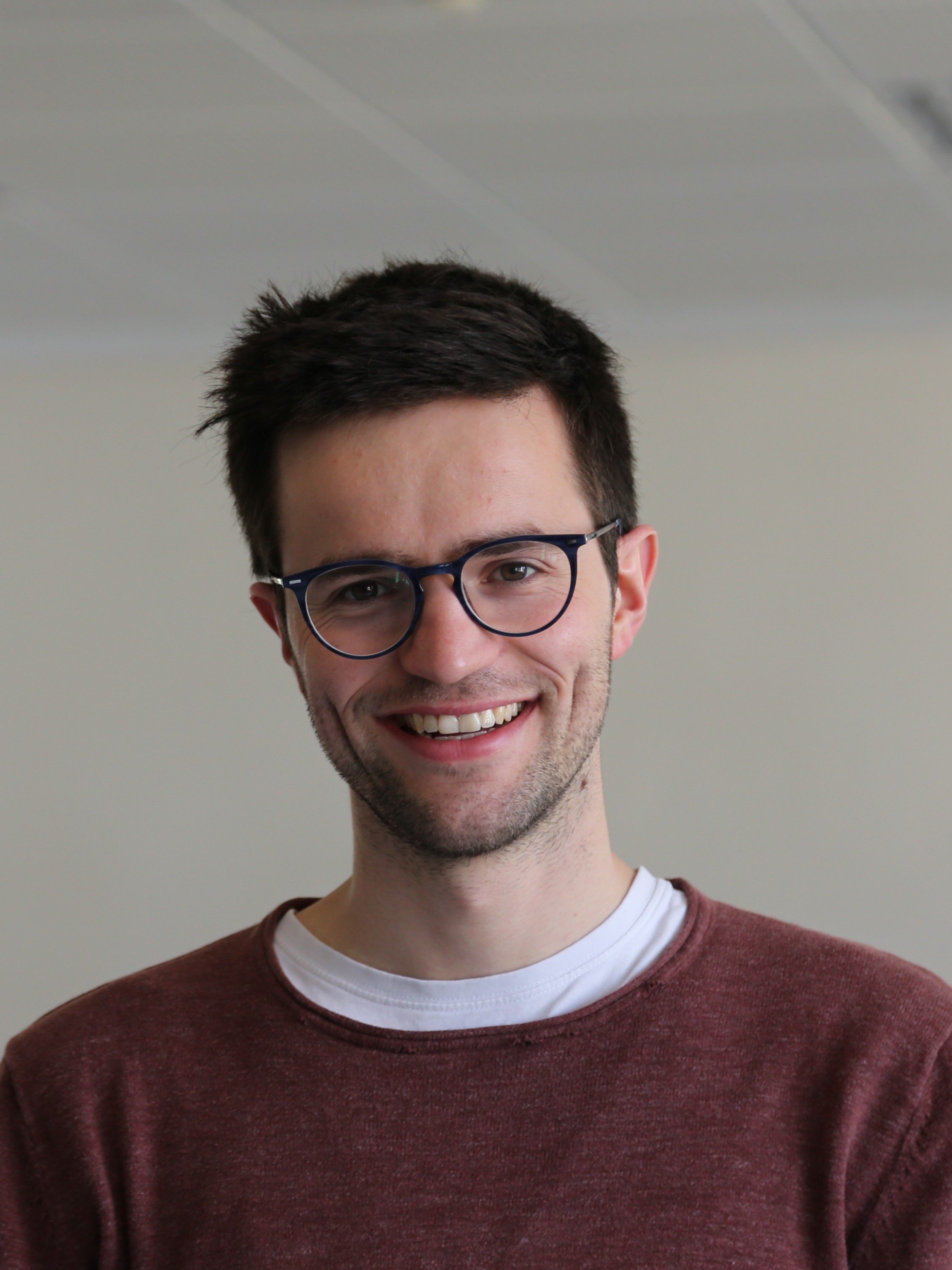}}]{Tobias M. Wolff} received a M.Sc. degree in Electrical Engineering from the Karlsruhe Institute of Technology and from the Grenoble Institute of Technology, both in 2021. He is currently a PhD student at the Institute of Automatic Control, Leibniz University Hannover, Germany. His main research interests are data- and learning-based estimation and biomedical applications.
\end{IEEEbiography}

\begin{IEEEbiography}[{\includegraphics[width=1in,height=1.25in,clip,keepaspectratio]{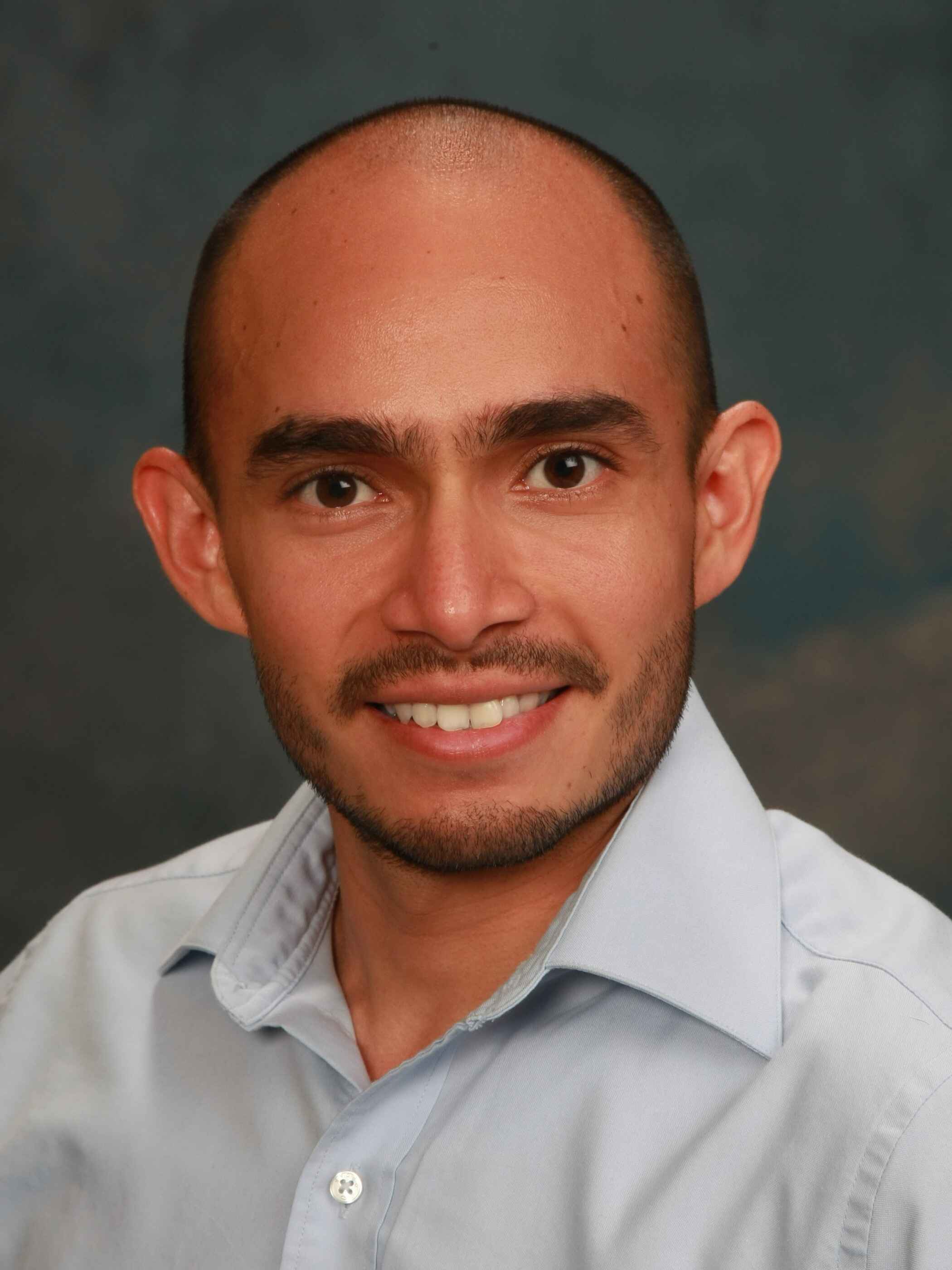}}]{Victor G. Lopez}  received his B.Sc. degree in Communications and Electronics Engineering from the Universidad Autonoma de Campeche, in Campeche, Mexico, in 2010, the M.Sc. degree in Electrical Engineering from the Research and Advanced Studies Center (Cinvestav), in Guadalajara, Mexico, in 2013, and his Ph.D. degree in Electrical Engineering from the University of Texas at Arlington, Texas, USA, in 2019. In 2015 Victor was a Lecturer at the Western Technological Institute of Superior Studies (ITESO) in Guadalajara, Mexico. From August 2019 to June 2020, he was a postdoctoral researcher at the University of Texas at Arlington Research Institute and an Adjunct Professor in the Electrical Engineering department at UTA. Victor is currently a postdoctoral researcher at the Institute of Automatic Control, Leibniz University Hannover, Germany. His research interest include cyber-physical systems, reinforcement learning, game theory, distributed control and robust control.
\end{IEEEbiography}

\begin{IEEEbiography}[{\includegraphics[width=1in,height=1.25in,clip,keepaspectratio]{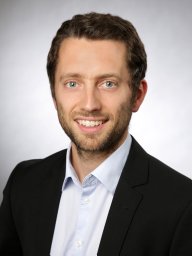}}]{Matthias A. Müller} (Senior Member, IEEE) received a Diploma degree in engineering cybernetics	from the University of Stuttgart, Germany, an M. Sc. in electrical and computer engineering	from the University of Illinois at Urbana-Champaign, US (both in 2009), and a Ph. D.	from the University of Stuttgart in 2014. Since 2019, he is Director of the Institute of Automatic Control and Full Professor at the Leibniz University Hannover, Germany. His research interests include nonlinear control and estimation, model predictive control, and data- and learning-based control, with application	in different fields including biomedical engineering and robotics. He has received various awards for his work, including the 2015 European Systems \& Control PhD Thesis Award, the inaugural Brockett-Willems Outstanding Paper Award for the best paper published in Systems \& Control Letters in the period 2014-2018, an ERC starting grant in 2020, the IEEE CSS George S. Axelby Outstanding Paper Award 2022, and the Journal of Process Control Paper Award 2023. He serves as an associate editor for Automatica and as an editor of the International Journal of Robust and Nonlinear Control.
\end{IEEEbiography}

\appendix

\subsection{Proof of Theorem~\ref{thm:MHE}}
\label{app:thm:proof}
The proof of Theorem~\ref{thm:MHE} is based on the developments shown in \cite{Schiller2023}. However, here we consider an MHE scheme based on learned dynamics (and not based on perfect model knowledge). Additionally, we consider a cost function with varying weights, which also requires some adaptations in the proof. Here, we mainly comment on the steps that are conceptually different from the proof in \cite{Schiller2023}, without describing the similar steps of the proof in all detail.

\textbf{Proof:}
The constraints in the MHE problem guarantee that the optimal estimated system trajectory (denoted by~$\hat{x}(j|t), u(j), \hat{w}(j|t), \hat{v}(j|t)$ for all $j \in \mathbb{I}_{[t- M_t, t-1]}$) satisfies the learned system dynamics~(\ref{def:learned_system}). In general, the unknown (true) system trajectory cannot necessarily be represented by the learned dynamics due to approximation errors. To represent the true system trajectory using the learned dynamics, we use the introduced auxiliary variables in~$\check{w}$~(\ref{def:auxiliary_w}) and in~$\check{v}$~(\ref{def:auxiliary_v}). We exploit that, due to~(\ref{eq:mean_in_MHE_h}), $m_{+,y}(h|d(j),D^{dt},Y^{dt}) - m_{+,y}(h|\hat{d}(j|t),D^{dt},Y^{dt}) = \hat{v}(j|t)-\check{v}(j)$ for all $j\in\mathbb{I}_{[t-M_t,t-1]}$. Furthermore, we apply inequality (\ref{ass:lyap:supply}) $M_t$ times along (i) the real and (ii) the optimal estimated trajectories, which yields
\begin{align*}
	W_\delta (\hat{x}&(t), x(t)) \nonumber \\
	& \overset{(\ref{ass:lyap:supply})}{\leq} \sum_{j =1 }^{M_t} \eta^{j-1} \Big(||\hat{w}(t-j|t) - \check{w}(t-j)||_{(\Sigma_x^{\max})^{-1}}^2\nonumber \\
	&\hspace{2cm}+ ||\hat{v}(t-j|t) - \check{v}(t-j)||_{(\Sigma_y^{\max})^{-1}}^2 \Big) \nonumber \\
	& \hspace{2cm}+\eta^{M_t}  W_\delta(\hat{x}(t-M_t|t), x(t-M_t)).
\end{align*}
Moreover, using (\ref{ass:lyap:bounds}), the triangle inequality, the Cauchy-Schwarz inequality, and Young's inequality, it holds
\begin{align*}
	&W_\delta (\hat{x}(t), x(t)) \nonumber\\
	&\leq2\eta^{M_t} 
	||\hat{x}(t-M_t) - x(t-M_t) ||_{(\Sigma_x^{\max})^{-1}}^2 \nonumber \\
	&+  \sum_{j =1 }^{M_t}2\eta^{j-1}\Big(||\check{w}(t-j)||_{(\Sigma_x^{\max})^{-1}}^2  + ||\check{v}(t-j)||_{(\Sigma_y^{\max})^{-1}}^2 \Big) \nonumber \\
	&+ 2 \eta^{M_t} ||\hat{x}(t-M_t|t) -  \hat{x}(t-M_t)||_{(\Sigma_x^{\max})^{-1}}^2\nonumber \\
	&+  \sum_{j = 1}^{M_t} 2\eta^{j-1}\Big(||\hat{w}(t-j|t)||_{(\Sigma_x^{\max})^{-1}}^2 + ||\hat{v}(t-j|t)||_{(\Sigma_y^{\max})^{-1}}^2 \Big)
\end{align*}
We define
\begin{align*}
	 &J_{\min}(\hat{x}(t-M_t|t), \hat{w}(\cdot|t), \hat{v}(\cdot|t),t) \\ &\coloneqq2 \eta^{M_t} ||\hat{x}(t-M_t|t) -  \hat{x}(t-M_t)||_{(\Sigma_x^{\max})^{-1}}^2\nonumber \\
	&+  \sum_{j = 1}^{M_t} 2\eta^{j-1}\Big(||\hat{w}(t-j|t)||_{(\Sigma_x^{\max})^{-1}}^2  + ||\hat{v}(t-j|t)||_{(\Sigma_y^{\max})^{-1}}^2 \Big) 
\end{align*}	
	
Note that $J_{\min}(\hat{x}(t-M_t|t), \hat{w}(\cdot|t), \hat{v}(\cdot|t),t)$ does not correspond to the optimal cost of problem (\ref{MHE_nom}). In fact, $J_{\min}(\hat{x}(t-M_t|t), \hat{w}(\cdot|t), \hat{v}(\cdot|t),t)$ corresponds to the cost of the optimal trajectory, when $(\Sigma_x^{\max})^{-1}$, $(\Sigma_y^{\max})^{-1}$ are considered in the cost function (\ref{eq:cost:function}) (but \textit{not} the variable matrices $(\Sigma_{x, \hat{d}(t-M_t-1|t-M_t)})^{-1}$, $(\Sigma_{x,\bar{d}(t-j|t)})^{-1}$, and $(\Sigma_{y,\bar{d}(t-j|t)})^{-1}$). 
Next, we upper bound $J_{\min}(\hat{x}(t-M_t|t),\hat{w}(\cdot|t),\hat{v}(\cdot|t),t)$ as follows
\begin{align}
	J_{\min}(\hat{x}(t-M_t|t),& \hat{w}(\cdot|t), \hat{v}(\cdot|t),t)  \nonumber \\
	&\leq 			J(\hat{x}(t-M_t|t), \hat{w}(\cdot|t), \hat{v}(\cdot|t),t) \label{eq:detec_optimal_cost} \\
	&\leq J(x(t-M_t), \check{w}(\cdot), \check{v}(\cdot),t) \nonumber
\end{align}	
where we recall that $J(\hat{x}(t-M_t|t), \hat{w}(\cdot|t), \hat{v}(\cdot|t),t) = J^\ast$ is the optimal value function of problem (\ref{MHE_nom}). The first inequality holds by~(\ref{rel:sigma_x}) and~(\ref{rel:sigma_y}), and the second is due to optimality (i.e., the true unknown system trajectory~$x(\cdot), \check{w}(\cdot), \check{v}(\cdot)$ is a feasible but in general suboptimal solution to problem~(\ref{MHE_nom})).
We consider these bounds and obtain
\begin{align}
	\label{eq:first:sigma_d}
	W_\delta &(\hat{x}(t), x(t)) \nonumber\\
	\leq& 2\eta^{M_t}||\hat{x}(t-M_t) - x(t-M_t) ||_{(\Sigma_x^{\max})^{-1}}^2 \nonumber \\
	&+\sum_{j =1 }^{M_t} 2\eta^{j-1}\Big(||\check{w}(t-j)||_{(\Sigma_x^{\max})^{-1}}^2+ ||\check{v}(t-j)||_{(\Sigma_y^{\max})^{-1}}^2\Big)\nonumber \\
	&+ 2\eta^{M_t}||\hat{x}(t-M_t) - x(t-M_t) ||_{(\Sigma_{x,\hat{d}(t-M_t-1|t-M_t)})^{-1}}^2 \nonumber \\
	&+\sum_{j =1 }^{M_t} 2\eta^{j-1}\Big(||\check{w}(t-j)||_{(\Sigma_{x,d(t-j)})^{-1}}^2 
	\nonumber \\
	&\hspace{2cm}+ ||\check{v}(t-j)||_{(\Sigma_{y,d(t-j)})^{-1}}^2\Big),
\end{align}
where $\Sigma_{x,d(t-j)}$ and $\Sigma_{y,d(t-j)}$ denote the uncertainty provided by the GPs when the true system trajectory is considered in the weighting matrices (\ref{def:weighting_matrix_x}) - (\ref{def:weighting_matrix_y}), initialized with $\Sigma_{x,d(t-M_t-1)} = 0$ (compare Footnote 4). Using~(\ref{rel:sigma_x}) and~(\ref{rel:sigma_y}), we obtain
\begin{align}
	W_\delta &(\hat{x}(t), x(t)) \nonumber\\
	&\leq    4\eta^{M_t}||\hat{x}(t-M_t) - x(t-M_t) ||_{(\Sigma_{x,\hat{d}(t-M_t-1|t-M_t)})^{-1}}^2 \nonumber \\
	&\hspace{0.2cm}+\sum_{j =1 }^{M_t} 4\eta^{j-1}\Big(||\check{w}(t-j)||_{(\Sigma_{x,d(t-j)})^{-1}}^2  
	\nonumber \\
	&\hspace{2cm}+ ||\check{v}(t-j)||_{(\Sigma_{y,d(t-j)})^{-1}}^2\Big) \label{eq:afterJmax}\\
	&\leq 4\lambda_{\max}\Big((\Sigma_{x}^{\min})^{-1},(\Sigma_{x}^{\max}+\varepsilon I)^{-1}\Big)\eta^{M_t}\nonumber \\
	&\hspace{0.2cm}\times	W_\delta (\hat{x}(t-M_t),x(t-M_t)) \nonumber \\
	&\hspace{0.2cm}+\sum_{j =1 }^{M_t} 4\eta^{j-1}\Big(||\check{w}(t-j)||_{(\Sigma_{x,d(t-j)})^{-1}}^2  
	\nonumber \\
	&\hspace{2cm}+ ||\check{v}(t-j)||_{(\Sigma_{y,d(t-j)})^{-1}}^2\Big). \nonumber
\end{align}
We choose $M$ large enough such that 
\begin{align}
	\label{eq:minimum_M}
	\mu^{M} \coloneqq 4\lambda_{\max}\Big((\Sigma_{x}^{\mathrm{min}})^{-1},(\Sigma_{x}^{\max}+\varepsilon I)^{-1}\Big)\eta^{M} <1
\end{align}
with $\mu \in [0,1)$, and obtain for all $t\in\mathbb{I}_{\geq M}$
\begin{align}
	W_\delta (\hat{x}(t), x(t)) \leq& \mu^{M} W_\delta (\hat{x}(t-M),x(t-M)) \nonumber \\
	&+\sum_{j =1 }^{M} 4\eta^{j-1}\Big(||\check{w}(t-j)||_{(\Sigma_{x,d(t-j)})^{-1}}^2  \nonumber \\
	&\hspace{1.2cm}+ ||\check{v}(t-j)||_{(\Sigma_{y,d(t-j)})^{-1}}^2\Big) \label{eq:M_step}.
\end{align}
We consider some time instant $t = \kappa M +l$ with $l \in \mathbb{I}_{[0, M-1]}$ and $\kappa \in \mathbb{I}_{\geq 0}$. Using~(\ref{eq:afterJmax}), we obtain
\begin{align}
	W_\delta(\hat{x}(l), x(l)) \leq& 4\eta^l ||\hat{x}(0) - x(0)||^2_{(\Sigma_{x,\hat{d}(-1|0)})^{-1}}  \nonumber \\
	&+ 4 \sum_{j =1 }^{l} \eta^{j-1} \Big( ||\check{w}(l-j)||_{(\Sigma_{x,d(l-j)})^{-1}}^2  \nonumber \\
	&\hspace{1.2cm}+ ||\check{v}(l-j)||_{(\Sigma_{y,d(l-j)})^{-1}}^2\Big) \label{eq:initial_FIE}.
\end{align}
The next step is to apply (\ref{eq:M_step}) $\kappa$ times. This means that we recursively iterate a contraction. Within each interval, we set the initial uncertainty to zero, compare the explanations in Subsection~\ref{subsec:design:weighting_matrices} (paragraph below (\ref{eq:derivatives})) and Footnote~4 in Section~\ref{sec:robust:stb}. With a slight abuse of notation, we do not specify this initialization in the following to improve the readability of the proof. After performing further steps which are similar to the proof of \cite[Cor. 1]{Schiller2023}, we obtain the following state estimation error bound
\begin{align*}
	&\|\hat{x}(t) - x(t)\|_{(\Sigma_{x}^{\max} + \varepsilon I)^{-1}} \\
	&\leq 2\sqrt{\mu}^t\|\hat{x}(0) - x(0)\|_{(\Sigma_{x,\hat{d}(-1|0)})^{-1}}\\
	&+\max_{q\in\mathbb{I}_{[0,t-1]}}\left\{\frac{2}{1-\sqrt[4]{\mu}}\sqrt[4]{\mu}^q\|\check{w}(t-q-1)\|_{(\Sigma_{x,d(t-q-1)})^{-1}}\right\}\nonumber \\ 
	&+\max_{q\in\mathbb{I}_{[0,t-1]}}\left\{\frac{2}{1-\sqrt[4]{\mu}}\sqrt[4]{\mu}^q\|\check{v}(t-q-1)\|_{(\Sigma_{y,d(t-q-1)})^{-1}}\right\}.
\end{align*}
Using 
\begin{align}
	a +\max_{i}\{b_i&\}+\max_i\{c_i\} \nonumber \\
	&\leq \max\{ 3a,\max_i\{3b_i\}, \max_i\{ 3c_i \}\} \label{eq:max}
\end{align}
for any $a,b_i,c_i \geq 0$ for $i = 1, \dots, t-1$, we obtain
\begin{align}
	&\|\hat{x}(t) - x(t)\|_{(\Sigma_{x}^{\max} + \varepsilon I)^{-1}} \nonumber  \\
	&\leq\ \max \Bigg\{ 6\sqrt{\mu}^t\|\hat{x}(0) -	x(0)\|_{(\Sigma_{x,\hat{d}(-1|0)})^{-1}}, \nonumber\\
	&\max_{q\in\mathbb{I}_{[0,t-1]}}\left\{\frac{6}{1-\sqrt[4]{\mu}}\sqrt[4]{\mu}^q\|\check{w}(t-q-1)\|_{(\Sigma_{x,d(t-q-1)})^{-1}}\right\},\nonumber \\ 
	&\max_{q\in\mathbb{I}_{[0,t-1]}}\left\{\frac{6}{1-\sqrt[4]{\mu}}\sqrt[4]{\mu}^q\|\check{v}(t-q-1)\|_{(\Sigma_{y,d(t-q-1)})^{-1}}\right\} \Bigg\}. \label{final_result}
\end{align}
We replace $\check{w}$ and $\check{v}$ according to~(\ref{def:auxiliary_w}) and~(\ref{def:auxiliary_v}), respectively. Then, we apply the triangle inequality and bound the difference between the functions $f$, $h$ and the posterior means $m_{+,x}$, $m_{+,y}$ by~(\ref{def:alpha1:max}) and~(\ref{def:alpha2:max}), respectively, which results in 
\begin{align}
	&\|\hat{x}(t) - x(t)\|_{(\Sigma_{x}^{\max} + \varepsilon I)^{-1}} \nonumber \\
	& \leq \max \bigg\{ 6\sqrt{\mu}^t\|\hat{x}(0) - x(0)\|_{(\Sigma_{x,\hat{d}(-1|0)})^{-1}}, \nonumber \\ 
	&\max_{q\in\mathbb{I}_{[0,t-1]}}\Big\{\frac{6\sqrt[4]{\mu}^q}{1-\sqrt[4]{\mu}}(\|w(t-q-1)\|_{(\Sigma_{x,d(t-q-1)})^{-1}} + \alpha_1^{\max})\Big\}, \nonumber \\
	&\max_{q\in\mathbb{I}_{[0,t-1]}}\Big\{\frac{6\sqrt[4]{\mu}^q}{1-\sqrt[4]{\mu}}(\|v(t-q-1)\|_{(\Sigma_{y,d(t-q-1)})^{-1}}  \nonumber \\
	&\hspace{5.7cm}+ \alpha_2^{\max})\Big\} \bigg\}. 
\end{align}
Using that $\max_{q\in\mathbb{I}_{[0,t-1]}}\sqrt[4]{\mu}^q =1$ and $a+b \leq \max\{2a,2b\}$ for any $a,b \geq 0$, we have
\begin{align}
	&\|\hat{x}(t) - x(t)\|_{(\Sigma_{x}^{\max} + \varepsilon I)^{-1}} \nonumber \\
	&\leq \max \Bigg\{ 6\sqrt{\mu}^t\|\hat{x}(0) - 	x(0)\|_{(\Sigma_{x,\hat{d}(-1|0)})^{-1}}, \nonumber \\ 
	&\max_{q\in\mathbb{I}_{[0,t-1]}}\left\{\frac{12}{1-\sqrt[4]{\mu}}\sqrt[4]{\mu}^q\|w(t-q-1)\|_{(\Sigma_{x,d(t-q-1)})^{-1}} \right\},\nonumber \\
	&\max_{q\in\mathbb{I}_{[0,t-1]}}\left\{\frac{12}{1-\sqrt[4]{\mu}}\sqrt[4]{\mu}^q\|v(t-q-1)\|_{(\Sigma_{y,d(t-q-1)})^{-1}} \right\}, \nonumber \\ 
	&\hspace{1cm} \frac{12}{1-\sqrt[4]{\mu}} \alpha_1^{\max},\frac{12}{1-\sqrt[4]{\mu}}\alpha_2^{\max}\Bigg\} \label{eq:nearly:final:result}. 
\end{align}
Finally, we use $\alpha_{\max}$ to bound the last two terms of~(\ref{eq:nearly:final:result}), which leads to the expression (\ref{thm:eq:expression}) of Theorem~\ref{thm:MHE}. \hfill$\blacksquare$

\subsection{Proof of Corollary~\ref{cor:prob:stab}}
\label{app:cor:proof}
We start with the result established in~(\ref{eq:nearly:final:result}).
We bound~$\alpha_1^{\max}$ as follows 
\begin{align}
	&\alpha_1^{\max} \leq \sqrt{\lambda_{\max}((\Sigma_{x}^{\min})^{-1})}  \times \nonumber \\
	&\sum_{i = 1}^{n} \max_{x \in \mathbb{X}, u \in \mathbb{U}} \big\{ || f_{i}(x,u) -m_{+, x_i}(f_i|d, D^{dt},X_i^{dt})|| \big \}, \label{eq:alpha:max1:prob} 
\end{align} 
and analogously for $\alpha_2^{\max}$ . \color{black}From here on, we apply \cite[Prop. 2]{fiedler2021practical} to probabilistically bound the difference between the true function components of $f$, $h$ and the corresponding posterior means. For the components\footnote{We assume for simplicity that the same value of $\delta$ has been chosen for all components of the functions $f$ and $h$.} of~$f$ and~$h$, this results in 
\begin{align*}
	&P\Big(||f_{i}(x,u) - m_{x_i}(f_i|d, D^{dt},X_i^{dt})||\\
	& \hspace{0.5cm} \leq B_{x_i}\sigma_{+,x_i}(f_i|d,D^d,X_i^d) + \eta_{N,x_i}(d), \forall x \in \mathbb{X}, u \in \mathbb{U}\Big)\\ 
	& \hspace{2cm} \geq 1- \delta  \hspace{0.5cm} i = 1, \dots, n \\
	&P\Big(||h_{j}(x,u) - m_{y_j}(h_j|d, D^{dt},Y_j^{dt})|| \\ 
	&\hspace{0.5cm} \leq B_{y_j}\sigma_{+,y_j}(h_j|d,D^d,Y_j^d) + \eta_{N,y_j}(d), \forall x \in \mathbb{X}, u \in \mathbb{U}\Big)\\ 
	& \hspace{2cm} \geq 1- \delta  \hspace{0.5cm} j = 1, \dots, p.
\end{align*}
The probability that all the components of $f$ jointly fulfill their bounds is lower bounded by $1-n\delta$ (the same holds for $h$ with probability $1-p\delta$) by applying the union bound. We replace the right hand side of (\ref{eq:alpha:max1:prob}) (and the analogous right-hand side of $\alpha_2^{\max}$) by these probabilistic bounds, i.e., 
\begin{align}
	P\big({\alpha}_1^{\max} &\leq \Delta_x^{\max}\big) \geq 1- n\delta \\
	P\big({\alpha}_2^{\max} &\leq \Delta_y^{\max}\big) \geq 1-p\delta.
\end{align}
Finally, the probability that both $\alpha_1^{\max} \leq \Delta_x^{\max}$ and $\alpha_2^{\max} \leq \Delta_y^{\max}$ hold jointly is lower bounded by $1-(n+p)\delta$. Using this bound in (\ref{eq:nearly:final:result}), we obtain the left-hand side of~(\ref{cor:prob:stab:expression}). \hfill $\blacksquare$


\subsection{Proof of~Theorem~\ref{thm:detec:orig:learned}}
\label{app:detec:proof}
Due to Assumption~\ref{ass:detec:orig}, the true system (\ref{def:system}) admits a $\delta$-IOSS Lyapunov function. We consider this~$\delta$-IOSS Lyapunov function, but replace the state transition function~$f$ by its approximation $m_{+,x}(x,u)$ from~(\ref{def:learned_system})
\begin{align*}
	W_\delta (&m_{+,x}(f|(x,u),D^{dt}, X^{dt}) + w,  \\
	&\hspace{2cm}m_{+,x}(f|(\tilde{x},u),D^{dt}, X^{dt})+ \tilde{w})\\
	=& W_\delta (f(x,u)+ w+ m_{+,x}(f|(x,u),D^{dt}, X^{dt}) - f(x,u), \nonumber \\
	&\hspace{0.5cm} f(\tilde{x},u)+ \tilde{w} +  m_{+,x}(f|(\tilde{x},u),D^{dt}, X^{dt})- f(\tilde{x},u)),
\end{align*}
where we added zero and then rearranged the terms. Now, we want to apply the dissipation inequality~(\ref{eq:IOSS_Lyap_2}), which holds for the original system. To this end, we introduce
\begin{align}
	w_{\mathrm{ld}} &\coloneqq w+ m_{+,x}(f|(x,u),D^{dt}, X^{dt}) - f(x,u) \label{def:w_ld}\\
	\tilde{w}_{\mathrm{ld}} &\coloneqq \tilde{w}+ m_{+,x}(f|(\tilde{x},u),D^{dt}, X^{dt}) - f(\tilde{x},u). \label{def:w_ld_tilde}
\end{align}
with $w_\mathrm{ld}, \tilde{w}_{\mathrm{ld}} \in \mathbb{W}$ by Assumption~\ref{ass:detec:sets} and $w, \tilde{w} \in \mathbb{W} \ominus \mathbb{E}_x$.
Applying inequality~(\ref{eq:IOSS_Lyap_2}) results in 
\begin{align}
	W_\delta (m_{+,x}&(f|(x,u),D^{dt}, X^{dt}) + w, \\
	&\hspace{1cm}m_{+,x}(f|(\tilde{x},u),D^{dt}, X^{dt})+ \tilde{w}) \nonumber \\
	\leq& \eta W_\delta (x,\tilde{x})+  \sigma_w\big(|| w_{\mathrm{ld}} - \tilde{w}_{\mathrm{ld}}||\big) \nonumber  \\ 
	&+\sigma_h\big(||h(x,u) - h(\tilde{x},u)||\big) 
\end{align}
We add zero on the right-hand side, rearrange the terms, and obtain
\begin{align*}
	W_\delta (m_{+,x}&(f|(x,u),D^{dt}, X^{dt}) + w, \\
	&\hspace{1cm}m_{+,x}(f|(\tilde{x},u),D^{dt}, X^{dt})+ \tilde{w}) \nonumber \\
	\leq& \tilde{\eta} W_\delta (x,\tilde{x}) + \sigma_w\big(|| w_{\mathrm{ld}} - \tilde{w}_{\mathrm{ld}}||\big) \\ 
	&+\sigma_h(||h(x,u) - h(\tilde{x},u)||) -(\tilde{\eta} - \eta) W_\delta (x,\tilde{x}). 
\end{align*}
Exploiting the properties of $\mathcal{K}$ functions, the definitions of $w_{\mathrm{ld}}$ and $\tilde{w}_{\mathrm{ld}}$ from (\ref{def:w_ld}) - (\ref{def:w_ld_tilde}), and adding zero in the argument of~$\sigma_h$ yields
\begin{align*}
	W_\delta &(m_{+,x}(f|(x,u),D^{dt}, X^{dt}) + w, \\
	&\hspace{2cm}m_{+,x}(f|(\tilde{x},u),D^{dt}, X^{dt})+ \tilde{w}) \nonumber \\
	& \leq \tilde{\eta} W_\delta (x,\tilde{x}) + \sigma_w(2|| (w -\tilde{w}||) \\
	&\quad+ \sigma_w(2||e_{x}(x,u) - e_{x}(\tilde{x},u)||) \\ 
	&\quad +\sigma_h(2||m_y (h|(x,u),D^{dt}, Y^{dt})  - m_y(h|(\tilde{x},u),D^{dt}, Y^{dt})||)  \\
	&\quad +\sigma_h(2||e_y(x,u) - e_y(\tilde{x},u) ||)  -(\tilde{\eta} - \eta) W_\delta (x,\tilde{x}),
\end{align*}
with $e_x$ and $e_y$ as defined in~(\ref{def:ex}) and (\ref{def:ey}), respectively. Due to Assumption~\ref{ass:detec:inequality}, we obtain
\begin{align*}
	W_\delta &(m_{+,x}(f|(x,u),D^{dt}, X^{dt}) + w, \\
	&\hspace{2cm}m_{+,x}(f|(\tilde{x},u),D^{dt}, X^{dt})+ \tilde{w}) \nonumber \\ 
	&\leq \tilde{\eta} W_\delta (x,\tilde{x}) +\sigma_w(2||w -\tilde{w}||) \nonumber\\
	&  +\sigma_h(2||m_y (h|(x,u),D^{dt}, Y^{dt})- m_y(h|(\tilde{x},u),D^{dt}, Y^{dt})||). 
\end{align*}
Since the upper and lower bounds as defined in~(\ref{eq:IOSS_Lyap_1}) still hold, the learned system admits a $\delta$-IOSS Lyapunov function~with $\tilde{\eta} \in [0,1)$, $\tilde{\sigma}_w(r) = \sigma_w(2r) \in \mathcal{K}$, $\tilde{\sigma}_h(r) = \sigma_h(2r) \in \mathcal{K}$.\hfill~$\blacksquare$

\end{document}